\documentclass{desyprocA4}
\usepackage{amssymb}
\newcommand\pubdate{\today}

\def\to{\rightarrow}

\def\bi{\begin{itemize}}
 \def\ei{\end{itemize}}

\def\c1p{C1^\prime}
\def\msq3{\overline{m}_{\tilde{q}}(3)}

\def\tst{\tilde t}
\def\ttau{\tilde \tau}

\def\tg{\tilde g}

\def\be{\begin{equation}}  
\def\ee{\end{equation}}  
\def\bea{\begin{eqnarray}}  
\def\eea{\end{eqnarray}}  
\def\tw{\tilde\chi}

\def\twpm{\tilde\chi^\pm}
\def\tz{\tilde\chi^0}

\newcommand{\eeto}    {\ensuremath{ {\, e}^+ {e}^- \to}}

\def\beq{\begin{equation}}
\def\eeq#1{\label{#1}\end{equation}}
\def\eeqn{\end{equation}}

\newenvironment{Eqnarray}%
   {\arraycolsep 0.14em\begin{eqnarray}}{\end{eqnarray}}
\def\beqa{\begin{Eqnarray}}
\def\eeqa#1{\label{#1}\end{Eqnarray}}
\def\eeqan{\end{Eqnarray}}

\newcommand{\ra}            {\ensuremath{ \rightarrow     }}

\begin{document}
%------------------------------------
%\title{Physics Beyond the Standard Model at the ILC}
%\title{The Case for US Support of the ILC Project:\\
\title{Physics Case for the ILC Project:\\
Perspective from Beyond the Standard Model}

%for single authors the superscripts are optional
\author{{\slshape Howard Baer$^1$, Mikael Berggren$^2$, Jenny List$^2$, Mihoko M. Nojiri$^{3,4}$,} \\ 
{\slshape Maxim Perelstein$^5$, Aaron Pierce$^6$,  Werner Porod$^7$, Tomohiko Tanabe$^8$}\\[1ex]
$^1$University of Oklahoma, Norman, OK 73019, USA\\
$^2$DESY, Notkestra{\ss}e 85, 22607 Hamburg, Germany\\
$^3$Theory Center, KEK, Tsukuba, Ibaraki 305-0801, Japan \\
$^4$Kavli IPMU, The University of Tokyo, Kashiwa, Chiba 277-8583, Japan \\
$^5$Cornell University, Ithaca, NY 14850, USA \\
$^6$University of Michigan, Ann Arbor, MI 48109, USA \\
$^7$University of W\"urzburg, 97074 W\"urzburg, Germany \\
$^8$ICEPP, The University of Tokyo, Bunkyo-ku, Tokyo 113-0033, Japan \\
}
% please enter the contribution ID for the DOI
%\contribID{xy}

% TO THE CONFERENCE EDITORS: 
% please update the following information      
% before sending the template to the authors
%\confID{   }  % if the conference is on Indico uncomment this line
%\desyproc{ \quad }
%\acronym{PLHC2010} % if you want the Acronym in the page footer uncomment this line
%\doi  % if there is an online version we will register DOIs

\maketitle

\pubdate

\begin{abstract}
The International Linear Collider (ILC) has recently proven its technical maturity with the
publication of a Technical Design Report, and there is a strong interest in Japan
to host such a machine. We summarize key aspects of the Beyond the Standard Model
physics case for the ILC in this contribution to the US High Energy Physics strategy
process. On top of the strong guaranteed physics case in the 
detailed exploration of the recently discovered Higgs boson, the top quark and electroweak
precision measurements, the ILC will offer 
unique opportunities which are complementary to the LHC program of the next decade.
Many of these opportunities have connections 
to the Cosmic and Intensity Frontiers, which we comment on in detail. We illustrate
the general picture with examples of how our world could turn out to be and what the ILC
would contribute in these cases, with an emphasis on value-added beyond the LHC. These comprise
examples from Supersymmetry including light Higgsinos, a comprehensive bottom-up coverage of NLSP-LSP
combinations for slepton, squark, chargino and neutralino NLSP, a $\ttau$-coannihilation 
dark matter scenario and bilinear R-parity violation as explanation for neutrino masses and mixing, 
as well as generic WIMP searches and Little Higgs models as non-SUSY examples.
\end{abstract}

% ****************************************************************************
%
\section{Introduction}
\label{sec:intro}
Experiments at the International Linear $e^+e^-$  Collider (ILC) may  be sensitive to new phenomena 
such as supersymmetric partners of known particles (SUSY), new heavy gauge bosons, extra spatial dimensions 
and particles connected with strongly-coupled theories of electroweak symmetry breaking~\cite{Intro2}. 
For accessible particles, ILC can yield substantial improvements over LHC measurements. 
In addition, ILC will have a qualitative advantage on signatures that have high backgrounds at LHC 
or are difficult to trigger on.  

In planning for future facilities relevant to exploring {\it physics beyond the Standard Model} (BSM), the proposed ILC
stands out as a mature and shovel-ready project which would provide unique features, making it 
complementary to the impending program of exploration by the LHC during the coming decade.
After more than twenty years of study, the ILC design has now achieved a state of
maturity  culminating recently with the publication of the Technical Design
Report \cite{Intro1}.
%The ILC has now achieved a state of
%maturity and readiness, culminating recently with the publication of the Technical Design
%Report \cite{Intro1}. 
Indeed, detailed simulations with realistic detector designs show that the ILC can achieve 
impressive precision \cite{Intro3}.  
%And just as the LHC experiments are now making more precise
%measurements than were originally predicted (as was also the case with the Tevatron,
%LEP and SLC experiments), the ILC experiments will bring qualitatively new capabilities
%and should similarly exceed the performance levels based on current simulations once
%data are in hand. 
The requirements of the ILC \cite{Intro4} include tunability between center-of-mass energies
of $\sqrt{s}=200$ and 500 GeV, with rapid changes in energy over a limited range for threshold
scans. 
Ultimately, expansion of the center-of-mass energy to $\sim 1$ TeV is envisioned.
The luminosity, which must exceed 10$^{34}$ cm$^{-2}$ s$^{-1}$ at 500 GeV, roughly scales
proportionally with center-of-mass collision energy. Highly polarized electrons ($>$80\%)
are specified, with polarized positrons desirable.   In this white paper, we will discuss how 
these capabilities allow for compelling explorations of physics at the weak scale and beyond.

Any discussion of weak scale physics should account for the enormously successful LHC runs at $\sqrt{s}=7$ and 8 TeV. 
Their crowning achievement has been the spectacular discovery of the long-awaited Higgs boson~\cite{Aad:2012tfa,Chatrchyan:2012ufa} 
with  $m_h = 125.5\pm 0.5$ GeV (ATLAS/CMS combined). 
But the discovery of the Higgs boson brings into sharp relief a well-known conundrum.  
To a good approximation, the newly discovered Higgs boson appears to be a fundamental scalar particle, 
and  scalar particles suffer from quadratic divergences to their mass squared.  In the case of the Higgs boson this leads to instability in the weak energy scale $\Lambda_\mathrm{weak}\sim 250$ GeV. 
Indeed, absent new physics up to the scale where gravity becomes strong, quantum fluctuations would drag the weak scale towards the scale $10^{19}$ GeV.
The marked discrepancy between these scales is the gauge hierarchy problem (GHP).  
A natural resolution suggests that the Higgs should be accompanied by other new particles not included in the Standard Model (SM), 
and furthermore these particles should be near the weak scale.

The nature of this BSM physics has been a subject of intense theoretical speculation 
for over three decades, and a number of theoretically attractive ideas have been proposed.
%Perhaps the most conservative 
One solution is to implement a quantum Fermi-Bose symmetry known as supersymmetry (SUSY). Supersymmetry softens 
the troublesome divergences in the Higgs mass to merely logarithmic ones, which do not place a significant upward pressure on the weak scale.  It does so while introducing a panoply of new states: squarks, sleptons, gauginos, and Higgsinos.
To naturally accommodate the weak scale, at least some of these superpartners
ought to have masses around the weak scale.  
Alternative approaches to solving the GHP include 
Little Higgs models \cite{LHModels} (which solve the GHP at one loop level, deferring it to $\sim 10$ TeV scale),
introduction of warped ``Randall-Sundrum" (RS) spacetime dimension with associated Kaluza-Klein excitations, 
composite Higgs models (often formulated in setups with additional extra dimensions of space such as RS models),
or postulating models with additional large spacetime dimensions. 
A qualitatively different spectrum of new matter states is predicted by each of these possibilities.

Strong new limits from the LHC do exist on new particle production.
For example, for squarks and gluinos \cite{SUSYAtlas,SUSYCMS}, limits in many cases push the masses of colored 
particles to values beyond a TeV.
In spite of these null results,  the complicated environment of proton collisions at the LHC does limit its power to hermetically search for new physics, 
and the current LHC data does not preclude a variety of well-motivated scenarios in which BSM particles are 
kinematically accessible at the ILC. These cases fall into several broad classes.

\begin{itemize}

\item BSM particles which only participate in electroweak interactions and are not charged under QCD strong force. 
Current LHC constraints on such particles are often much less stringent than those on colored particles, 
due to small production cross sections and large backgrounds. 
A well-motivated example is chargino pair production which is predicted in SUSY models. 
Even in cases where LHC production cross sections are large, the indistinct nature of the signal is difficult
to extract from large $t\bar{t}$ and $W^+W^-$ backgrounds.
On the other hand, chargino pair production would be easily seen at an ILC provided $\sqrt{s}>2m_\mathrm{\twpm}$
due to the much more constrained kinematics. 

\item BSM particles where decays do not release large amounts of visible energy. 
For example, current bounds on mixed chargino-neutralino production at the LHC~\cite{ATLASElectroweakinos} vanish if the mass splitting between 
the produced states and the lightest superpartner is less than 50 GeV. 
Difficulty exploring such regions persists even at very high luminosity \cite{ATLASHighLumSUSY}.  
Well-motivated examples of such compressed spectra exist. 
For instance, naturalness considerations in SUSY models suggest the existence of light Higgsinos with masses as light as $\sim 100-200$ GeV. 
And while Higgsino pair production may occur at sufficient rates at LHC, their highly compressed 
spectrum leads to only soft visible energy release in their decays, making detection exceedingly difficult. 
In the clean environment of an ILC, Higgsino pair production is straightforward to observe, provided that $\sqrt{s}>2m_\mathrm{Higgsino}$.

\item A third class pertains to the case where BSM particles decay to purely hadronic channels. An example-- 
again from SUSY models-- occurs when the lightest neutralinos can be pair produced, but where these decay 
to purely hadronic states via $R$-parity violating operators. In the case of LHC 
these may be buried beneath substantial QCD backgrounds whilst at an ILC they could be easily observed.

\item A fourth class occurs when the BSM particles lie somewhat beyond the TeV scale. 
In this case, neither LHC nor ILC would have enough energy to produce them directly. 
However, quantum loop and other virtual effects in electroweak precision observables 
may allow an ILC to glean knowledge of their existence indirectly, perhaps determining the
energy scale where the new matter states could be directly detectable. 
(For such studies, we defer to the Precision Electroweak Group report.)
\end{itemize}

In these scenarios, the ILC will be able to discover new physics missed by the LHC. Of course, there is also the exciting possibility that LHC makes a new physics discovery in the upcoming high energy/high luminosity runs.
In this case, the role of the ILC would be to 
%use its 1. clean scattering environment, 2. low background rates, 
%3. tunable beam energies and 4. tunable beam polarization to {\it 
precisely characterize any new matter states which are
accessible.

The Japanese government and HEP community have taken the leading role in offering to act as host country for such a machine.
The European community has voiced strong support for ILC construction in Japan. 
Support from the US would provide critical momentum to help get this project started.
%The time is ripe for the United States to jump on board, and provide the critical momentum to get this project started.   
%Recently, the Japanese government has expressed a desire to host the ILC, and
%international negotiations are underway.
In a staged approach, beginning at a center-of-mass energy of 250 GeV, a physics program would start with precision measurements of
the Higgs branching ratios and properties. Raising the energy to 500 GeV would move
to precision measurements of top quark properties well beyond those possible at the
LHC.   Should there be accessible new particles such as supersymmetric
partners of gauge bosons and leptons, the ILC is the only place where they can be studied
in full detail. Extension of the ILC to 1 TeV is straightforward, with lengthened linac tunnels and additional cryomodules, building on
the original ILC sources, damping rings, final focus and interaction regions, and beam
dumps.

%While the ILC is envisioned to  function initially 
%as a Higgs factory, if naturalness arguments from supersymmetry are correct, then it might well turn out to be 
%in addition a Higgsino factory!

This brief white paper is intended to highlight a variety of the unique and essential capabilities that an ILC would
bring to the search for New Physics. 
We will conclude that:
\begin{itemize}
\item Even after the initial phase of LHC running, ILC remains a discovery machine. It is likely to remain so
even into the next decade. A machine with true hermetic sensitivity to new physics up to a TeV will have powerful implications for whether or not the electroweak scale is, in fact, natural.
%Either ILC discovers light Higgsinos or rules out SUSY EW naturalness!
\item If new physics is discovered at LHC, ILC with tunable energy, polarized beams, low background, 
    theoretically well-understood interactions, precision beam energy and capacity for threshold scans would
    be a precision microscope for determining detailed properties of all low lying states.
\item Precision measurements may allow for extrapolation to much higher mass scales, {\it e.g.} tests of unification.
\item ILC has a unique role to play in dark matter physics, including 
1. the possible observation of direct  WIMP (weakly interacting massive particle) pair 
production via the recoil of an initial state radiation (ISR) photon and 
2. precision measurements of new physics properties which would constrain and test dark matter production 
in the early universe along with
providing particle physics input to direct and indirect WIMP search experiments.
\item Even if new matter states turn out to be beyond ILC reach (and beyond the scope of this report which is devoted to
direct production of new matter states), precision measurements sensitive to virtual quantum effects 
can also provide critical information. 
\end{itemize}
For these reasons, from the standpoint of BSM physics, we conclude that there is a strong case for the ILC. 
Its rich program of new physics exploration-- especially in light of the recent Higgs discovery-- 
offers an exciting frontier facility for the US HEP community in the coming decades.

%
% *****************************************************************************

% ****************************************************************************
%
\section{Connection to Cosmic Frontier}
\label{sec:cosmo}
During the past several decades, a very compelling and simple scenario has
emerged to explain the presence of dark matter (DM) in the universe with an abundance roughly
five times that of baryonic matter. The {\it WIMP miracle} scenario posits that 
weakly interacting massive particles would be in thermal equilibrium with the cosmic
plasma at very high temperatures $T> m_\mathrm{WIMP}$. As the universe expands and cools, 
the WIMP particles would freeze out of thermal equilibrium, locking in a relic abundance
that depends inversely on the thermally-averaged WIMP (co)-annihilation 
cross section~\cite{Lee:1977ua}.
The WIMP ``miracle'' occurs in that a weak strength annihilation cross section gives
roughly the measured relic abundance provided the WIMP mass is of the order of the 
weak scale~\cite{Baltz:2006fm}.  The lightest neutralino of SUSY models has been touted as a 
%protypical
WIMP candidate~\cite{Goldberg:1983nd,Ellis:1983ew,Jungman:1995df}.
The lightest $T$-parity-odd particle in Little Higgs theories and the lightest
Kaluza-Klein (KK) excitation in extra-dimensional models preserving KK-parity also can serve
as possible WIMP candidates. 

%And while the WIMP miracle scenario is both simple and engaging, it is now clear that
%it suffers from several problems, at least in the case of SUSY theories.
The WIMP miracle is both simple and engaging.  
And while the WIMP miracle gets the relic density right to within orders of magnitude, 
detailed agreement with the measured thermal relic density is not guaranteed for just any weak scale candidate.  
Indeed, in SUSY models the correct thermal abundance is often achieved in regions of parameter space that require 
special relationships between the underlying parameters. 
For example, bino-like WIMPs typically annihilate away inefficiently, and are thus overabundant.  
They can successfully be depleted by co-annihilation processes, {\it e.g.} with a stau~\cite{EllisCo},  
but this requires a near degeneracy (within 5\%) between the stau and the neutralino.  
Alternatively, the DM could annihilate through a resonance, but this requires $m_\mathrm{DM} \approx m_\mathrm{res}/2$. 
Such considerations present an opportunity:  the ILC has the precision to verify whether these special relationships 
are indeed satisfied~\cite{Baltz:2006fm}, and thus weigh in on whether the universe does obey a 
thermal history back to temperatures of order 100 GeV. 
Such an exploration would represent a roughly four order-of-magnitude increase in knowledge of physics 
at temperature scales beyond those probed by Big Bang Nucleosynthesis (BBN).

% These include:
%\begin{itemize}
% \item While the WIMP miracle gets the relic density right to within orders of magnitude, 
%in practice bino-like WIMPs are overproduced while wino- or Higgsino-like WIMPs are 
%underproduced. Special mechanisms-- WIMP co-annihilation, tuning of bino-Higgsino-wino mixing, also known as  tempering) or resonance annihilation-- are required to match the predicted relic abudance with astrophysical measurements.  Any of these mechanisms require 

It is also plausible that the WIMP miracle is not the whole story, indeed:
\begin{itemize}

\item In SUSY theories, gravitinos may also play a major role. In might be the case that the gravitino
forms the bulk of dark matter, instead of a WIMP. Gravitinos can be produced thermally in the early universe, 
or via sparticle cascade decays. In addition, if gravitinos are heavier than the lightest 
SUSY particle (LSP), then their decays may augment the neutralino abundance; however, 
their production and (Planck suppressed) decays are tightly constrained by BBN.

\item The presence of light moduli-fields from string theory may influence the relic abundance, either
by decaying into DM particles (augmenting the standard abundance) or by decaying into SM particles
(thereby diluting all relics present at the time of decay).

\item The strong $CP$ problem cries out for a solution, and so far the most compelling is that
of the semi-visible Peccei-Quinn (PQ) axion. The axion can also serve as dark matter particle. 
Furthermore, in SUSY theories the axion is accompanied by a spin-$1/2$ axino and a spin-0 saxion. These particles can also
augment or dilute the dark matter abundance depending on various PQ parameters.
\end{itemize}
Thus, the physics associated with dark matter production may be much more complicated than just the simple picture 
provided by the standard WIMP miracle scenario.

The ILC can play a unique role in providing critical information needed to identify the  dark matter
particle(s) and their production in the early universe. In the case of theories like SUSY, Little Higgs models
or extra dimensional models with KK dark matter, the high precision measurement of low lying 
new particle masses, branching fractions, spin and other quantum numbers will provide strong particle physics
constraints. Such measurements would offer terrestrial laboratory input to calculations of dark matter 
production rates in the early universe. If the inferred relic density lies above or below the standard thermal estimate, 
then evidence may be gleaned for non-standard processes such as the presence of axions.  
Characterization of dark matter properties will also provide input to direct and indirect  WIMP detection calculations.

In particular, ILC may be able to produce DM or DM-related particles ({\it e.g.} the next-to-lightest 
SUSY particle, or NLSP) either directly or via cascade decays. In the former case, $e^+e^-\rightarrow\chi\chi\gamma$
production would allow laboratory tagging of invisible particles which might make up the dark matter.
We will give more details about this possibility in 
Sec.~\ref{subsec:WIMPs}.
In the latter case, kinematic constraints from identifying heavier DM-related particles could allow for
a direct measurement of the DM particle's mass {\it e.g.} via the jet-jet energy distribution 
from $e^+e^-\rightarrow \chi^+\chi^-\rightarrow (\ell^+\nu_{\ell}\chi)+(q\bar{q}^\prime\chi )$. In case the
observed relic density comes about with the help of co-annihilation processes, it is not enough to measure
the properties of the WIMP, but in addition precise information on the mass and mixing of its co-annihilation
partner is needed. We illustrate this case with an example in Sec.~\ref{subsec:stc}.

Finally, ILC may be able to indirectly weigh in on the issue of baryogenesis. While electroweak baryogenesis
seems highly constrained in light of recent LHC results, it does require rather light top squarks.
If these are nearly mass degenerate with the LSP, then they may well elude LHC searches,  but should still be accessible to ILC. 
Alternatively, the Affleck-Dine baryogenesis mechanism requires the presence of baryon- and/or lepton-number 
carrying scalar fields. Such fields are aplenty in SUSY models, and their discovery and characterization at ILC 
would lend credence, albeit indirectly, to this approach. 
%Finally, the most generic theories of leptogenesis require large 
%re-heat temperatures after inflation $T_R>10^9$ GeV. But such a large $T_R$ tends to overproduce gravitinos.
%Discovery of superparticles and possibly gravitinos and characterization of thei%r properties at an ILC
%would lead to important new input to these compelling ideas for the origin of matter.

%
% ****************************************************************************

% ****************************************************************************
%
\section{Connection to Intensity Frontier}
\label{sec:intensity}
%
% ****************************************************************************
%\begin{itemize}
% \item Itensity Frontier:
%   \begin{itemize}
%       \item Neutrino mixing \lra\ bRPV SUSY
%       \item rare SUSY decays, eg stop \ra\ charm
%   \end{itemize}
%\end{itemize}
% ****************************************************************************
In the last fifteen years another branch of particle physics has seen impressive
advances and extension of our knowledge: flavor physics. 
%After the first experimental
%appearance of the so-called atmospheric neutrino puzzle in 1998 
Starting with the compelling data on atmospheric neutrinos in 1998, it has now been firmly established
that neutrinos are massive, and the different flavors are strongly
mixed. However, neutrino masses are much smaller than the masses of the electrically charged
fermions and flavor mixing in the leptonic sector differs considerably from that in the
quark sector. These experimental findings, along with the observed hierarchies in
the masses of the charged leptons and quarks form the so-called flavor puzzle.
Even though we can parametrize our understanding, 
%using 22 parameters (12 masses, 6 mixing angles
%and 4 CP-phases) 
we do not know the underlying principles leading to the experimentally
observed patterns.
In the last decade,  in addition to substantial progress in the neutrino sector,
B-factories have considerably extended our knowledge of the quark sector. 
Nowadays LHC-- where for example a first measurement of $BR(B_s \to \mu^+ \mu^-)$ has 
recently been performed by the LHCb collaboration-- is also contributing to this endeavor.

From the theoretical side, various models have been proposed which aim to explain
the observed flavor structures, and several of them predict new particles at the TeV
scale. Often these flavor models are combined with supersymmetry to ensure the
relative stability of the various scales involved. Typical examples
are supersymmetric versions of the Froggatt-Nielsen mechanism as discussed
for example in Ref.~\cite{Ross:2004qn}. In such cases traces of this additional
sector are left in the supersymmetry breaking parameters.  This, in turn, leads to
additional contributions to flavor and CP violating observables in low
energy experiments. From data in the Kaon and B-meson sectors
we know that these additional contributions are either sub-dominant or they conspire
for an unknown reason
%, which however could be related to the mechanism
%responsible for flavour structures, 
such that all relevant observables in 
the quark sector seem to be SM-like. Such a ``conspiracy'' can occur  
because one has to sum over all possible virtual contributions which usually 
appear only at higher order.
However, such a conspiracy cannot happen for the decays of supersymmetric particles 
(or those of any other sector of a BSM model related to quarks) and thus the measurement of
various decay properties can provide a direct window into the flavor sector. 
While at the LHC squarks and gluinos may well be produced copiously, 
it remains experimentally very challenging to determine the branching ratios
of the individual particles \cite{delAguila:2008iz}. 
However, theoretical arguments prefer that at least one of the stops
has mass in the range of a few hundred GeV and thus might be accessible to the ILC. 
Detailed information on its branching ratios would give
important information on flavor structures beyond the ones present
in the SM \cite{Weiglein:2004hn}.

Several models have been proposed to explain the smallness of neutrino masses.
At low energies they usually lead to the so-called Weinberg operator $(LH)^2$
giving Majorana masses to light neutrinos after electroweak symmetry breaking.
The most popular among them is the seesaw mechanism which comes in three varieties
depending on the details of how this operator is generated: type I in case of gauge singlet
fermions (usually the right-handed neutrinos), type II in case of an $SU(2)
_L$ triplet
Higgs boson and type III in case of $SU(2)_L$ triplet fermions. These additional
particles are usually too heavy to be produced directly in collider experiments.
In supersymmetric models they leave imprints in the RGE evolution of
the mass parameters~\cite{Baer:2000hx} and give rise to additional flavor structures which are linked
to the underlying mechanism for generating neutrino masses. These flavor structures
induce flavor violating decays of sleptons which can be studied  at the ILC
\cite{Deppisch:2003wt}.

Beside the usual mechanisms to generate neutrino masses, supersymmetry offers an 
additional possibility: breaking of $R$-parity in the lepton sector. The simplest
model is the one where only bilinear terms are present in the superpotential as well
as the corresponding terms in the soft SUSY sector. As $R$-parity is broken,
the lightest supersymmetric particle (LSP) is not stable anymore but decays. 
The six parameters explaining neutrino data are then also those responsible for the decay
properties of the LSP: ratios of decay branching ratios are proportional
to neutrino mixing angles, {\it e.g.}
$BR(\tilde \chi^0_1 \to W \mu)/BR(\tilde \chi^0_1 \to W \tau) \simeq \tan^2\theta_\mathrm{atm}$ or
$BR(\tilde \chi^0_1 \to \nu \mu \tau)/BR(\tilde \chi^0_1 \to \nu e \tau) \simeq \tan^2\theta_\mathrm{sol}$
\cite{Porod:2000hv} where
$\theta_\mathrm{atm}$ and $\theta_\mathrm{sol}$ are the atmospheric and solar neutrino mixing angles. 
We will illustrate this possibility with a dedicated simulation study in section~\ref{subsec:bRPV}.
Moreover, the smallness of the neutrino masses
also implies that the lifetime of the LSP is measurable at the ILC in a large part of the parameter
space. For completeness we note that the existence of such correlations does not depend
on the nature of the LSP-- {\it e.g.} whether it is a neutralino or a chargino or a slepton-- but only
the concrete form of these correlations \cite{Hirsch:2003fe}. 

%
% ****************************************************************************

% ****************************************************************************
%
\section{ILC Stories}
\label{sec:stories}
%
% *****************************************************************************

While there is a compelling case that new structure is needed in the laws of physics
at the $100-1000$ GeV scale, exactly what that structure might be is one of the 
great mysteries of our time. Theoretical guidance allows one to extend the 
Standard Model to include new ideas such as supersymmetry, dark matter, grand unification, 
see-saw neutrinos, extra dimensions, string theory, and many others. But whether, or how, any of these ideas would
manifest themselves at the weak scale is open to speculation; the truth will only be gleaned 
by a program of detailed experimental tests.
The ILC-- with its uniquely clean and flexible experimental environment -- 
is prepared for almost any possibility. In this white paper, we focus on scenarios in which new particles 
appear within kinematic reach of the ILC. Here, we present seven such scenarios -- each a particular story of how nature 
might be constructed at the weak scale, and what role the ILC could play in revealing the new physics. 

%It is, of course, very likely that the truth does not correspond exactly to any of the scenarios considered here: as
%R. P. Feynman once said: ``the imagination of nature is far, far greater than that of man''. However, the breadth of 

\subsection{Coming to terms with electroweak naturalness} 
\label{subsec:natsusy}

Supersymmetric theories provide an elegant solution to the gauge hierarchy problem. 
However, a lack of SUSY signals at LHC8 combined with the rather
large value of $m_h\sim 125$ GeV seemingly exacerbates what has come to be known as the
Little Hierarchy problem (LHP): why is there such a discrepancy between the electroweak
scale, typified by $m_Z=91.2$ GeV and $m_h=125.5$ GeV, and the superpartner scale, 
which in the case of gluinos and squarks, seems to be at the TeV-or-beyond scale.
Phrased differently, one might wonder why, if superpartners are at the $>1$ TeV scale,
the $Z$ mass is just $91.2$ GeV instead of also at the $>1$ TeV scale?

In the Minimal Supersymmetric Standard Model (MSSM), an answer can be extracted from the electroweak scalar potential minimization condition
which relates $m_Z^2$ to the SUSY breaking parameters and the superpotential 
Higgsino mass $\mu$:
\begin{equation}
\frac{m_Z^2}{2} =
\frac{m_{H_d}^2 + \Sigma_d^d -(m_{H_u}^2+\Sigma_u^u)\tan^2\beta}{\tan^2\beta -1} -\mu^2 \;,
\label{eq:loopmin}
\end{equation}
where $\Sigma_u^u$ and $\Sigma_d^d$ include a variety of 
radiative corrections~\cite{Baer:2012up,Baer:2012cf}.
To {\it naturally} obtain a $Z$ mass of 91.2 GeV, one expects each
contribution to the right-hand-side of Eq. \ref{eq:loopmin} to also be $\sim m_Z^2/2$:
{\it i.e.} there are no large uncorrelated contributions to the $Z$-mass.

To allow for electroweak naturalness, {\it e.g.} requiring no worse than cancellations at the 
$\sim 3\%$ level, then it is necessary that 
(a) $|\mu | \sim 100-300$ GeV,
(b) $m_{H_u}^2$ is driven to only small negative values under RG evolution and
(c) the top squarks $\tst_1$ and $\tst_2$ are highly mixed with masses
$m_{\tst_1}\sim 1-2$ TeV and $m_{\tst_2}\sim 2-4$ TeV.
The large mixing softens the top squark radiative corrections while at the same time lifting 
$m_h$ up to $\sim 125$ GeV.

When these conditions are met, then one may allow for a {\it natural} Little Hierarchy characterized by
\begin{itemize}
\item $m_{\mathrm{Higgsino}}\sim m_Z\sim m_h $
\item top squarks which enter Eq.~\ref{eq:loopmin} at one-loop level and gluinos
should live in the $1-5$ TeV regime and
\item first/second generation squarks and sleptons which enter  Eq.~\ref{eq:loopmin}
at two-loop level can exist at the $10-20$ TeV regime, which allows for at least
a partial solution to the SUSY flavor and $CP$ problems.
\end{itemize}

The main implication of this picture-- dubbed radiatively-driven natural supersymmetry (RNS) 
because the soft term $m_{H_u}^2$ is radiatively driven to small negative values at the 
electroweak scale~\cite{Baer:2012up,Baer:2012cf}-- 
is that there should exist four light physical
Higgsinos $\tz_1$, $\tz_2$ and $\tw_1^\pm$ with mass $\sim 100-300$ GeV (the lighter the better) where
$\tz_1$ is the LSP which is dominantly Higgsino-like (albeit with a 
non-negligible gaugino component). Due to the compressed spectrum 
amongst the various Higgsino states (typically a 10-20 GeV mass gap in models with
gaugino mass unification), their three-body decays yield only tiny visible energy
release, making them very difficult to detect at LHC.
On the other hand, the light Higgsinos should be easily detected at an ILC
provided that $\sqrt{s}>2|\mu |$.

\begin{figure}[htb]
  \begin{center}
\includegraphics[width=0.425\textwidth]{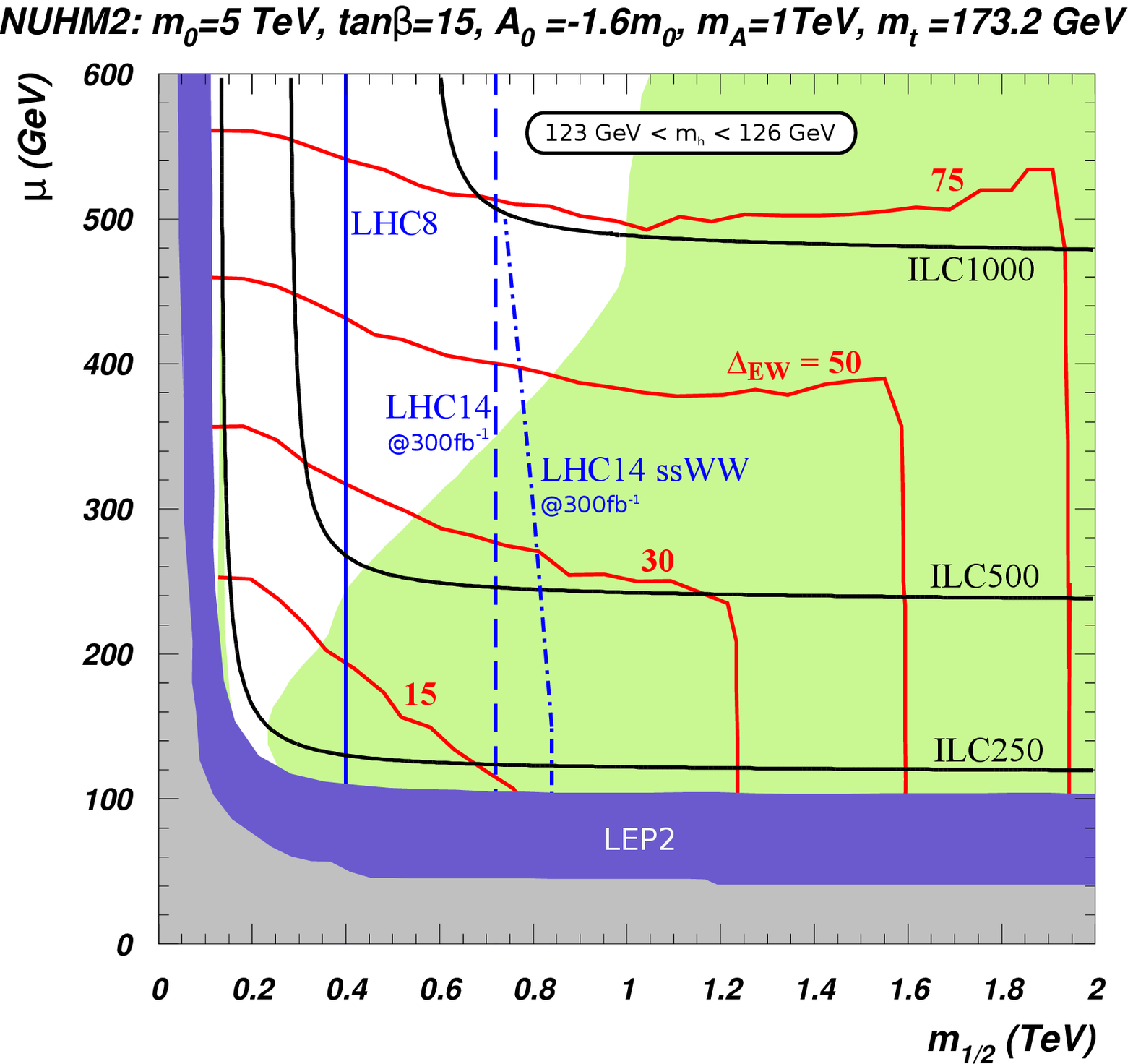}
\hspace{0.2cm}
\includegraphics[width=0.485\textwidth]{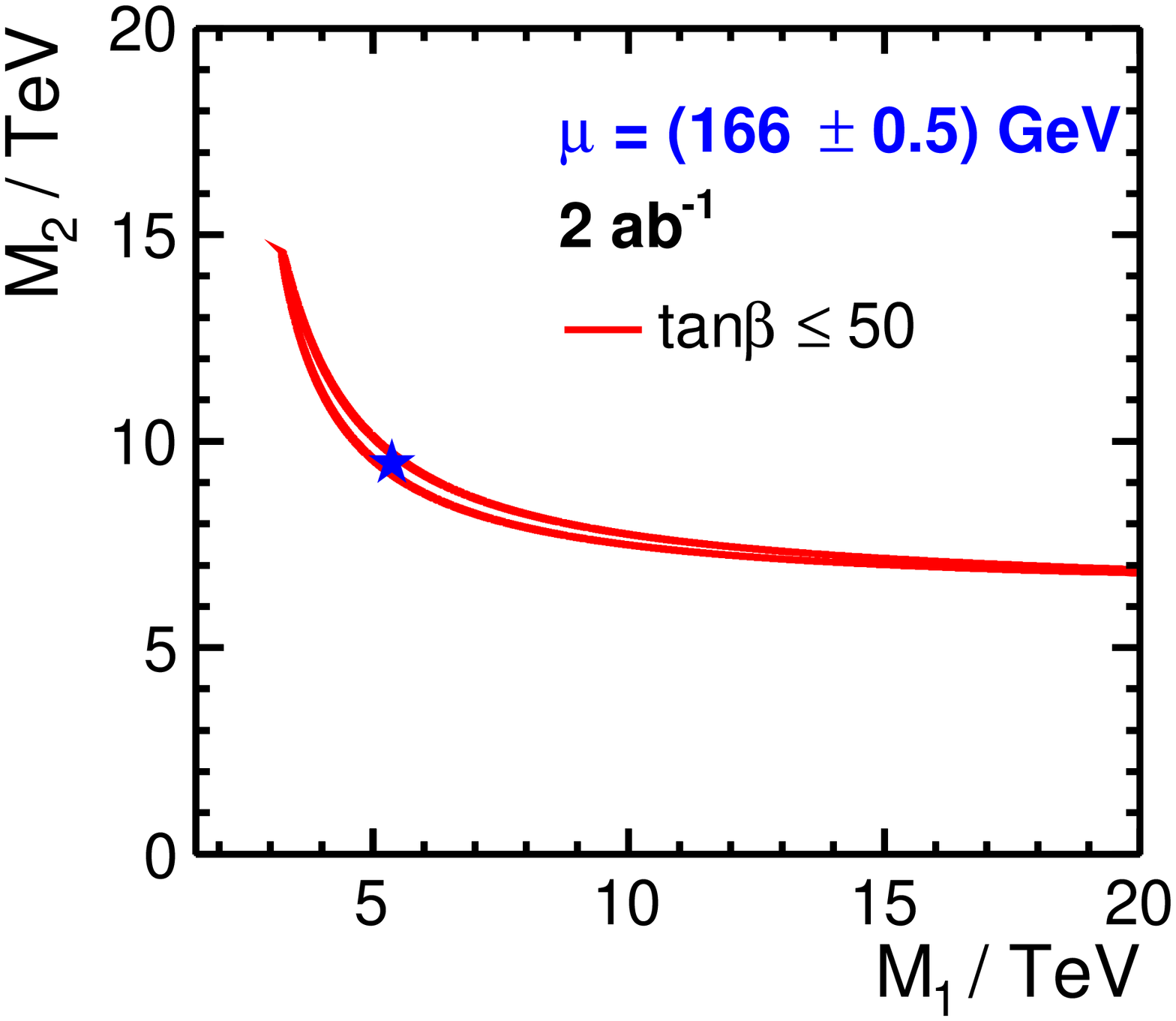}
  \end{center}
  \caption{Left: Contours of fine-tuning $\Delta_\mathrm{EW}=15,\ 30,\ 50$, and 75 in the $m_{1/2}$ vs. $\mu$ plane for
the RNS model with parameters as shown (from Ref.~\cite{Baer:2013faa}). 
The blue vertical lines show the current reach of LHC8 and the projected reach of 
LHC14 with 300 fb$^{-1}$ via gluino pair searches (dashed line) and same-sign dibosons (dot-dashed).
The reach of ILC with $\sqrt{s}=250$, 500 and 1000 TeV is also shown.
The green-shaded region has a thermal Higgsino relic abundance
$\Omega_{\tilde h}h^2\le 0.12$. 
Right: Parameter determination in the $M_2$ vs $M_1$ plane in the benchmark Higgsino scenario.
The star indicates the input values of $M_2$ and $M_1$, the $\Delta \chi^2 = 1$ contour of a 
fit to expected ILC measurements is displayed in red. From the same fit, $\mu$ is obtained with
sub-GeV precision. From Ref.~\cite{bib:higgsino_paper}.
}
\label{fig:rnsplane}
\end{figure}

The situation can be illustrated within the $\mu\ vs.\ m_{1/2}$ plane in the RNS model~\cite{Baer:2013faa}, 
where we also take GUT scale matter scalar masses $m_0=5$ TeV, $\tan\beta =15$, $A_0=-1.6 m_0$ and $m_A=1$ TeV. 
From the left panel of Fig.~\ref{fig:rnsplane}, it can be seen that LHC8 has explored 
$m_{1/2}\lesssim 0.4$ TeV via the search for $\tg\tg$ production. 
The calculated LHC14 reach with 300$^{-1}$ fb for $\tg\tg$
production~\cite{Baer:2012vr} and for same-sign diboson production~\cite{Baer:2013yha} extends to 
$m_{1/2}\sim 0.7-0.8$ TeV (corresponding to a reach in $m_{\tg}\sim 1.8-2.1$ TeV). 
The naturalness contours of $\Delta_\mathrm{EW} =30$ ({\it i.e.} $\Delta_\mathrm{EW}^{-1}\sim 3\%$ fine-tuning) 
extend well beyond LHC14 reach all the way to $m_{1/2}\sim 1.2$ TeV. 
However, ILC600 can probe the entire parameter space with $\Delta_\mathrm{EW}< 30$,
thus either discovering Higgsinos or ruling out SUSY electroweak naturalness.
%
%%%%%%%%%%%%%%%%%%%%%%%%%%%%%%%%%%%%%%%%%%%%%%%%%%%%%%%%%%%%%%%%%%%%%%%%%%%%%%%%%%%%%%%
%%%%%%%%%%%%%%%%%%%%%%%%%%%%%%%%%%%%%%%%%%%%%%%%%%%%%%%%%%%%%%%%%%%%%%%%%%%%%%%%%%%%%%

This is true independently of the size of the  mass splittings between the Higgsinos.
The clean environment of the ILC allows to resolve and measure mass differences 
even in the sub-GeV regime. Beam polarization can not only be employed to enhance the signal,
but also to verify that $s$-channel $Z$ (or $Z/\gamma$) exchange is the only production 
mechanism as expected for Higgsinos.

The achievable precision at the ILC has recently been studied for a benchmark scenario with 
$\mu = 166$~GeV, assuming ILC TDR~\cite{Intro1} machine parameters and detector 
performance~\cite{Intro3}. The gaugino mass parameters are as large as $M_1 = 5.4$~TeV 
and $M_2 = 9.5$~TeV, resulting in a mass difference of only 
$\Delta(M_{\twpm_1} - M_{\tz_1}) = 770$~MeV~\cite{bib:lc2013_sert,bib:higgsino_paper}.
Even in such a challenging situation, all states can be fully resolved. 
$M_{\twpm_1}$ and $M_{\tz_2}$ can be determined to a few GeV, while the statistical 
uncertainty on the mass difference is a few tens of MeV. The polarized cross-sections 
can be measured at the percent level.

Already with an integrated luminosity of $500$~fb$^{-1}$ at $\sqrt{s}=500$~GeV and 
$P(e^+,e^-) = (+30\%,-80\%)$, these measurements are precise enough to determine $\mu$ 
at the percent level, and to put lower limits in the few TeV-range on $M_1$ and 
$M_2$~\cite{bib:lc2013_rolbiecki,bib:higgsino_paper}. With higher luminosity, $M_1$ and 
$M_2$ can actually be constrained to a narrow region, as displayed in the right part of
Fig.~\ref{fig:rnsplane}. The star indicates the input values of $M_2$ and $M_1$, the 
$\Delta \chi^2 = 1$ contour of a 
fit to expected ILC measurements is displayed in red. From the same fit, $\mu$ is obtained with
sub-GeV precision. This shows that already with $\sqrt{s} = 500$~GeV, the ILC can constrain
parameters in the multi-TeV range and thus predict the physical masses of the heavier states.

In summary, light Higgsinos are an important discovery opportunity for the ILC, even if no new physics 
is found at LHC14. Especially for $m_{1/2} \gtrsim 1$~TeV, only the ILC can systematically 
explore the regime of natural supersymmetry independently of any assumptions on the top 
squark or the gluino mass, and either exclude the notion of natural SUSY or become a 
{\it Higgsino factory} in addition to being a Higgs factory.

\subsection{At the ILC, SUSY {\itshape is} simplified}
\label{subsec:simplified}

At lepton colliders, one has the possibility
to search for SUSY in a model-independent way~\cite{Tsukamoto:1993gt}:
The cornerstone of SUSY is that {\it sparticles couple as particles}.
This is independent of the mechanism responsible for SUSY breaking.
In particular, the coupling to the $Z$ boson is known, which
implies that the cross section for any
$\eeto$ sparticle-antisparticle pair process in the $s$-channel
is determined by the kinematics alone, {\it i.e.} by $\sqrt{s}$ and the mass of the
sparticle.
Furthermore, by definition there is {\it one} LSP,
and {\it one} NLSP.
Cosmological arguments shows that if the LSP is stable, it 
must be neutral and weakly 
interacting.
The NLSP, on the other hand, could be any sparticle: a slepton,
an electroweakino, or even a squark.
However, there are only a {\it finite} number of sparticles.

\begin{figure}[htb]
  \begin{center}
  \includegraphics[width=0.425\textwidth]{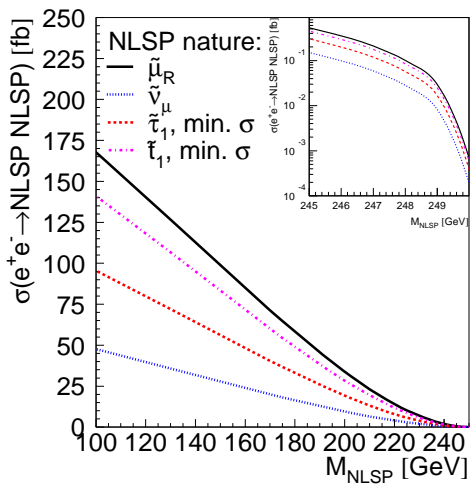}
  \end{center}
  \caption{NLSP production cross sections as a function
of M$_\mathrm{NLSP}$ at  $\sqrt{s}$ = 500 GeV (best beam polarization) for a few NLSP
hypotheses. The insert shows the last few GeV before the
kinematic limit, on a log scale.
}
\label{fig:simple1}
\end{figure}

While an arbitrary sparticle in the spectrum of any
SUSY model typically would decay through cascades
of other, lighter sparticles, the NLSP only has {\it one} decay mode,
namely to the LSP and the SM partner of the NLSP.
Therefore, studying NLSP production and decay can be regarded
as a ``simplified model without simplification'': Any SUSY 
model will have such a process.

Putting these observations together,
one realizes that by
systematically searching for signals for all possible NLSP's,
the entire space of models that are within kinematic reach of the ILC can be covered.

\subsubsection*{Closing the loopholes}

At the ILC, a systematic search for the NLSP is possible 
{\it without leaving loopholes}, covering even the cases that may be very difficult to test at the LHC.

In the case of a very small mass difference between the
LSP and the NLSP - less than a few GeV -
the clean environment at the ILC nevertheless allows
for a good detection efficiency.
If $\sqrt{s}$ is much larger than the threshold for the
NLSP-pair production, the NLSPs themselves will be
highly boosted in the detector frame, and most of the
spectrum of the decay products will be easily detected.
In this case,  
the precise knowledge of
the initial state at the ILC is of paramount importance to recognize the
signal, by the slight discrepancy in energy, momentum and acolinearity
between signal and background from pair production of the NLSP's SM
partner.
In the case the threshold is not much below  $\sqrt{s}$,
the background to fight is $\gamma\gamma \to f\bar{f}$
where the $\gamma$'s are virtual ones radiated off the
beam-electrons. The beam-electrons themselves are deflected
so little that they leave the detector undetected through
the outgoing beam-pipes.
Under the clean conditions at the ILC,
this background can be kept under control by demanding that
there is a visible ISR photon accompanying the soft NLSP decay 
products. If such an ISR is present in a $\gamma\gamma$ event,
the beam-remnant will also be detected, and the event can be rejected.

If the LSP is unstable due to R-parity violation, the ILC reach would be better or equal to the
$R$-conserving case, both for long-lived and short-lived
LSP's and whether the LSP is charged or neutral.

Also in the case of an NLSP which is a
mass-state mixed between the hyper-charge states,
the procedure is viable.
One will have one more parameter - the mixing angle.
However, as the couplings to the Z of both states are
known from the SUSY principle, so is the coupling with
any mixed state. There will then be one mixing-angle that
represents a possible ``worst case'', which allows to
determine the reach whatever the mixing is - namely the
reach in this ``worst case''. 
%% Only a neutralino NLSP might pose a problem in that
%% the mixing-structure is more complex, and usually the
%% there will be correlations between the nature of the
%% LSP and the NLSP. Therefore, both cross-sections and decay
%% products are influenced by the mixing, and it will be hard to
%% determine the reach in the  $M_\mathrm{NLSP} - M_\mathrm{LSP}$ plane. However,
%% the reachable cross-section can be evaluated, 
%% and can be confronted with any given model.

Finally, the case of ``several'' NLSPs-- {\it i.e.} a group
of near-degenerate sparticles-- can be disentangled due to 
the possibility to precisely choose the beam energy at the ILC.
This will make it possible to study the ``real'' NLSP below the threshold of
its nearby partner.

\begin{figure}[ht]
  \begin{center}
\includegraphics[width=0.425\textwidth]{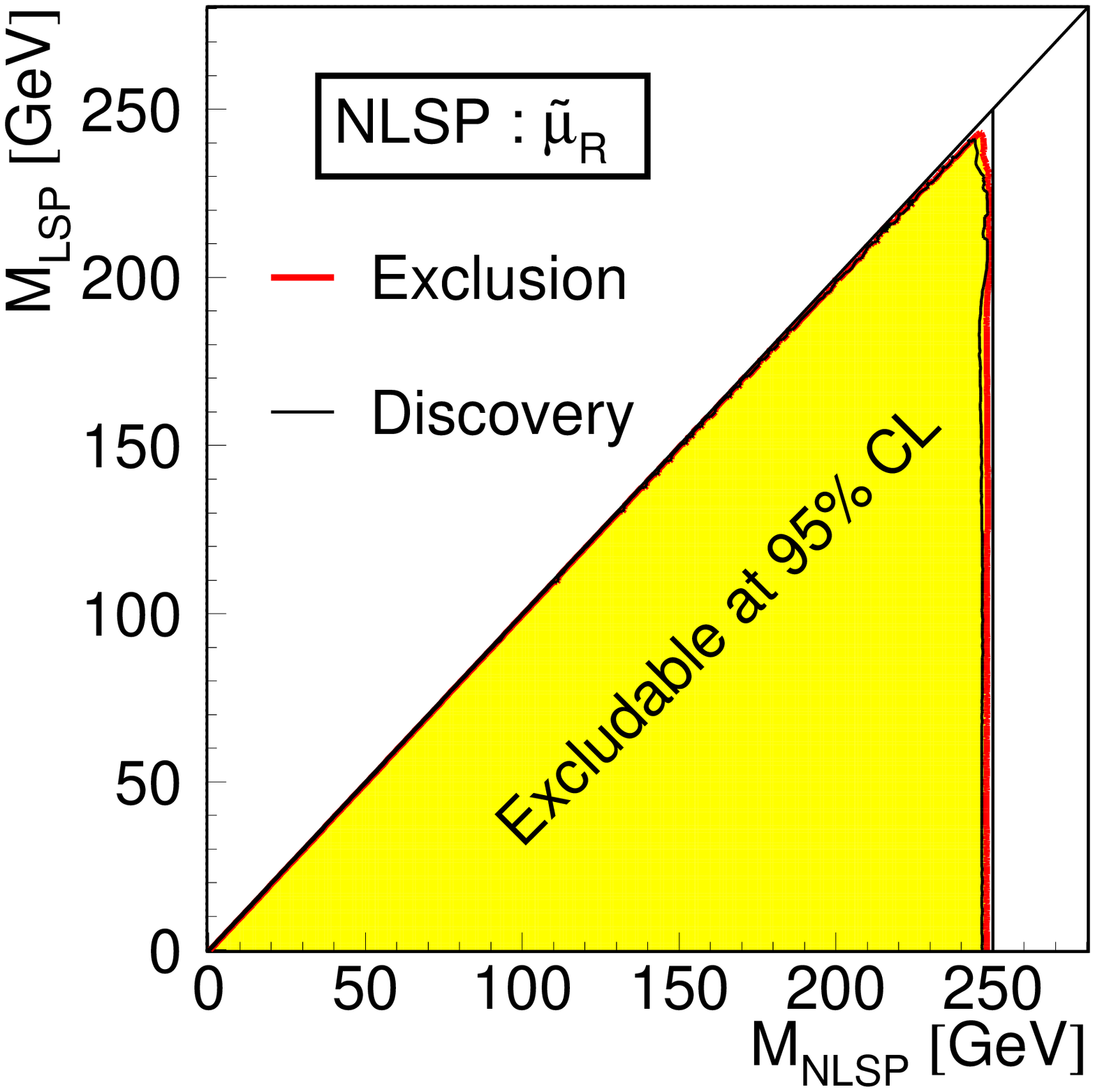}
\hspace{0.2cm}
\includegraphics[width=0.425\textwidth]{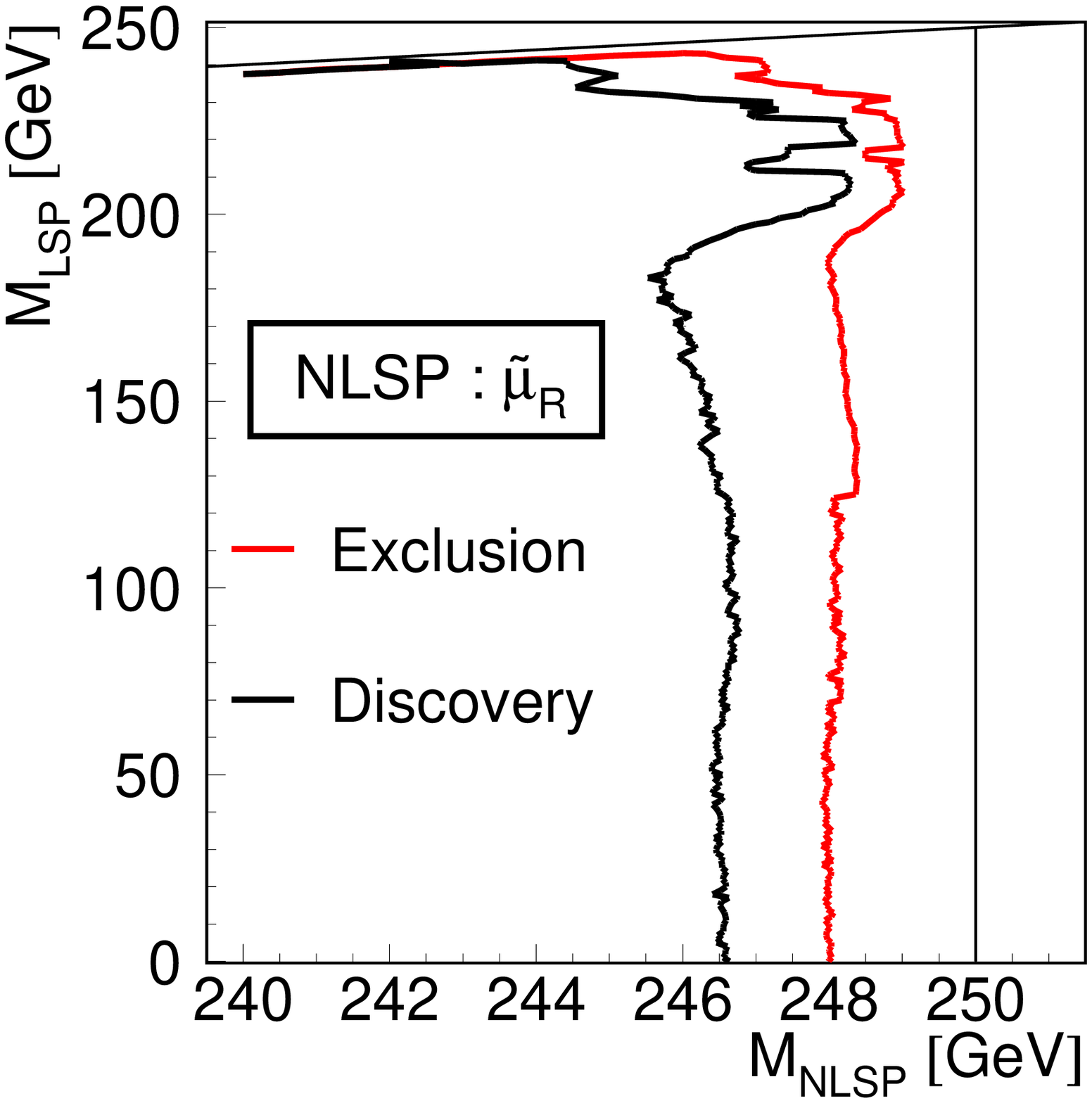}
  \end{center}
  \caption{Discovery reach for  a $\tilde{\mu}_R$ NLSP after 
collecting 500 fb$^{-1}$
at $\sqrt{s}$ = 500 GeV. Left: full scale, Right: zoom to last few GeV before the
kinematic limit.
}
\label{fig:simple2}
\end{figure}

\subsubsection*{The strategy}

At an $e^+e^-$-collider, the following typical features of NLSP production
and decay can be exploited: 
missing energy and momentum, high acolinearity,
expected particle or jet flavor identification, as well as invariant di-jet/di-lepton
mass conditions, optionally using constrained kinematic fitting.
A very powerful feature due to the known initial state at the ILC
is that the kinematic edges of the detected systems
can be precisely calculated at any point in the  $M_\mathrm{NLSP} - M_\mathrm{LSP}$ plane.
In particular, close to kinematic limit where the width
of the decay product spectrum is quite small, this feature allows for
an almost background-free signal with high efficiency.

To estimate the background at each point,
correctly normalized samples of 
events from all SM processes are generated, passed
through detector simulation, and the analysis chain.
If the number of observed events 
passing the selection criteria for a given
NLSP nature exceeds the
expected background passing the same cuts by more than  5$\sigma$,
one can claim {\it discovery} of the NLSP.
If, on the other hand, the observed number does not
exceed the expected background by more than 2$\sigma$,
{\it exclusion} can be claimed.
As this procedure is performed for {\it every possible NLSP},
it will constitute a complete and model-independent search for SUSY.

In order to estimate what the expected {\it Discovery Reach}
or  {\it Exclusion Reach}
of the experiment is, it is enough
to simulate the signal for each possible NLSP
in a fine grid in the  $M_\mathrm{NLSP} - M_\mathrm{LSP}$ at the given $\sqrt{s}$,
calculate the production cross section from the
SUSY principle and kinematics,
and confront it to the relevant selection criteria.

%Combined with the integrated luminosity and the 
%production cross-section, calculated from the
%SUSY-principle and kinematics, this gives the number of expected events.
%Together with simulated background
%this allows to determine if
%the point is within {\it Discovery Reach} - expected significance
%$< 5\sigma$ - or  within {\it Exclusion Reach} - expected significance
%between 2 and 5 $\sigma$.

\begin{figure}[hb]
  \begin{center}
\includegraphics[width=0.425\textwidth]{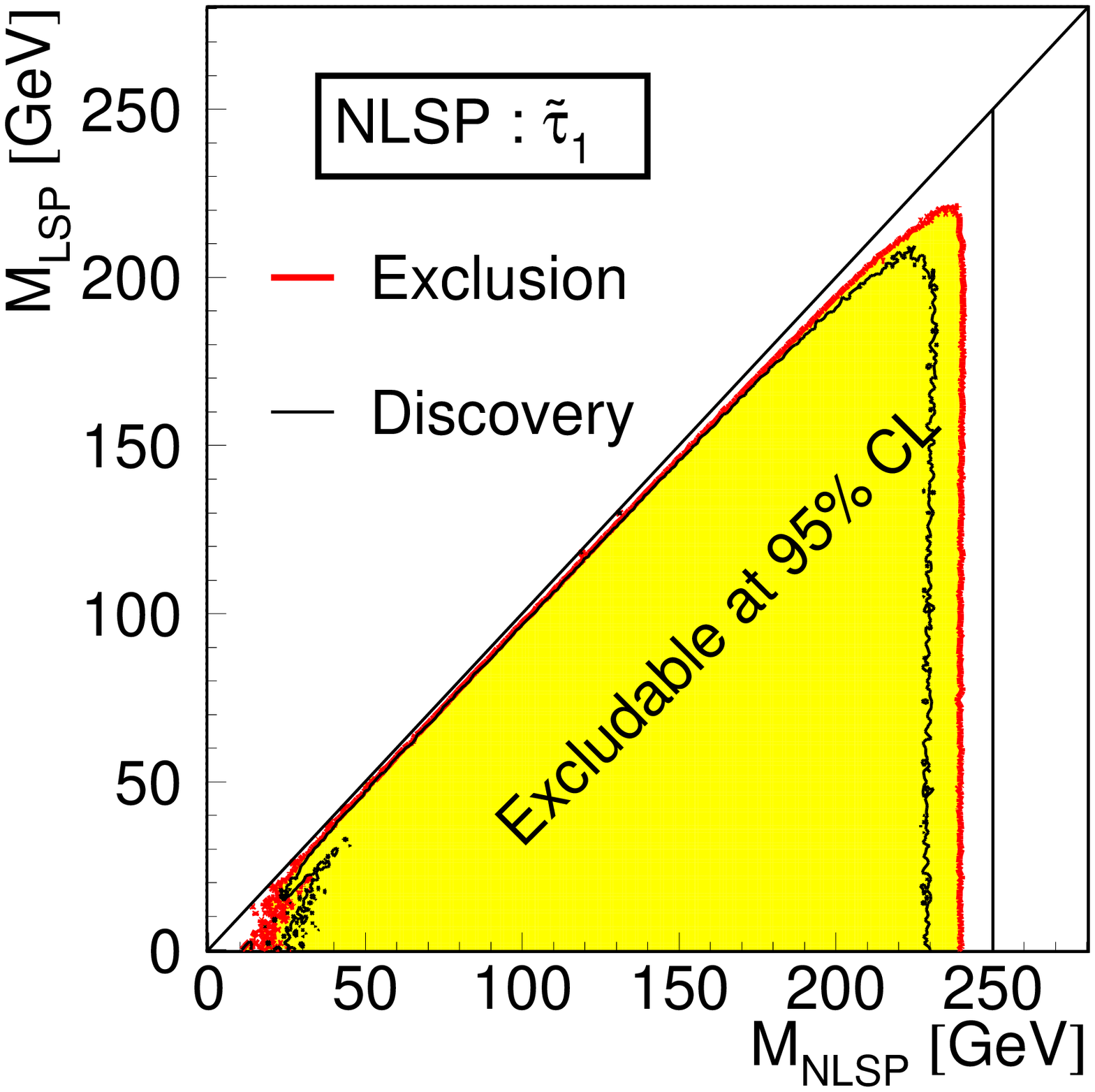}
\hspace{0.2cm}
\includegraphics[width=0.425\textwidth]{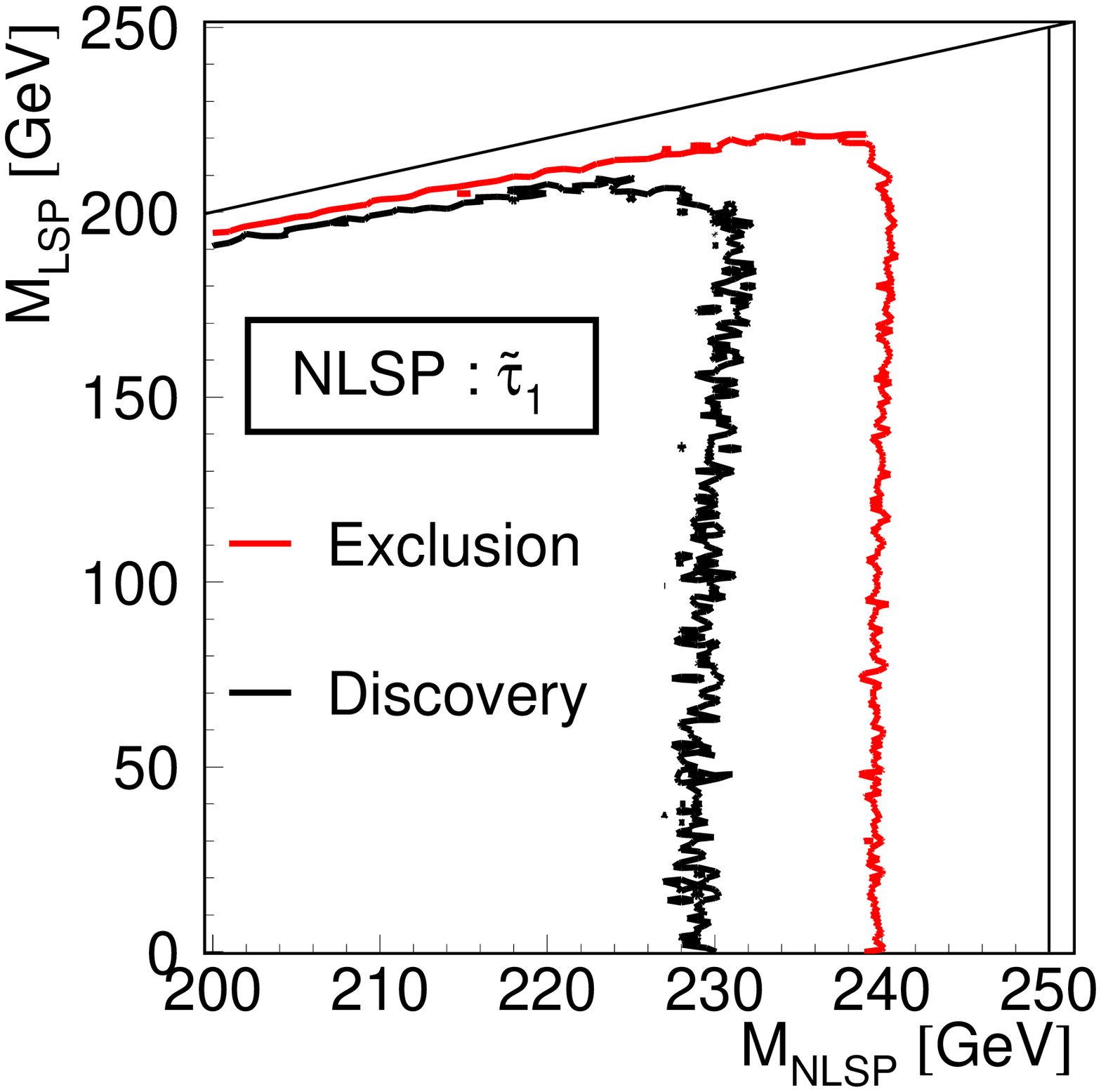}
  \end{center}
  \caption{Discovery reach for  a $\tilde{\tau}_1$ NLSP after 
collecting 500 fb$^{-1}$
at $\sqrt{s}$ = 500 GeV. Left: full scale, Right: zoom of the region close to the
kinematic limit.
}
\label{fig:simple3}
\end{figure}

% %%onefig_begin
% In Fig.~\ref{fig:simple}, the cross-section at $\sqrt{s}$ = 500 GeV
% as a function of $M_\mathrm{NLSP}$ is shown in the left panel
% for a selection of NLSP-candidates. The right panel shows as example
% the $5\sigma$ discovery and $2\sigma$ exclusion
% reach for a $\tilde{\mu}_R$ NLSP after collecting 500 fb$^{-1}$
% at $\sqrt{s}$ = 500 GeV.
% \begin{figure}[htb]
%   \begin{center}
%   \includegraphics[width=0.425\textwidth]{nlsp_xsect_w_snu.eps}
% \hspace{0.2cm}
% \includegraphics[width=0.425\textwidth]{smu_nlsp_reach.eps}
%   \end{center}
%   \caption{Left: NLSP production cross-sections as a function
% of M$_\mathrm{NLSP}$ at  $\sqrt{s}$ = 500 GeV (best beam-polarization) for a few NLSP
% hypotheses. The insert shows the last few GeV before the
% kinematic limit, on a log-scale.
% Right: Discovery-reach for  a $\tilde{\mu}_R$ NLSP after 
% collecting 500 fb$^{-1}$
% at $\sqrt{s}$ = 500 GeV. 
% }
% \label{fig:simple}
% \end{figure}
% %%onefig_end
%twofigs_begin
In Fig.~\ref{fig:simple1}, the cross section at $\sqrt{s}$ = 500 GeV
as a function of $M_\mathrm{NLSP}$ is shown 
for a selection of NLSP candidates,
and in  Fig.~\ref{fig:simple2} (Fig.~\ref{fig:simple3}), as example,
the $5\sigma$ discovery and $2\sigma$ exclusion
reach for a $\tilde{\mu}_R$  NLSP  ($\tilde{\tau}_1$ NLSP) after collecting 500 fb$^{-1}$
at $\sqrt{s}$ = 500 GeV.
%%In  
%%Figs.~\ref{fig:simple2} and~\ref{fig:simple3},
%%the $5\sigma$ discovery and $2\sigma$ exclusion
%%reach for $\tilde{\mu}_R$ and $\tilde{\tau}_1$  NLSPs after collecting 500 fb$^{-1}$
%%at $\sqrt{s}$ = 500 GeV.
%%
%%twofigs_end

%\subsection{LHC and ILC complementarity: Non-simplified SUSY (Mikael/Jenny)}
\subsection{LHC and ILC complementarity: SUSY is complex!}
\label{subsec:stc}
\def\XIPM#1{\ensuremath{ \tilde{\chi}^{\pm}_#1}}
\def\XI0#1{\ensuremath{ \tilde{\chi}^0_#1}}
\def\stau#1{\ensuremath{ \tilde{\tau}_#1}}
\def\smu#1{\ensuremath{ \tilde{\mu}_#1}}
\def\sel#1{\ensuremath{ \tilde{e}_#1}}

In full SUSY models, the higher states of the spectrum can have many decay modes leading to
potentially long decay chains~\cite{Baer:1988kx}. This means that the simplified approach in general does not 
apply beyond the direct NLSP production case discussed in the previous section, which
renders the interpretation of exclusion limits formulated in
the simplified approach non-trivial. Furthermore, also many production channels may be open, 
making SUSY the most serious background to itself.

\begin{figure}[htb]
%  \begin{center}
\includegraphics[width=0.52\textwidth]{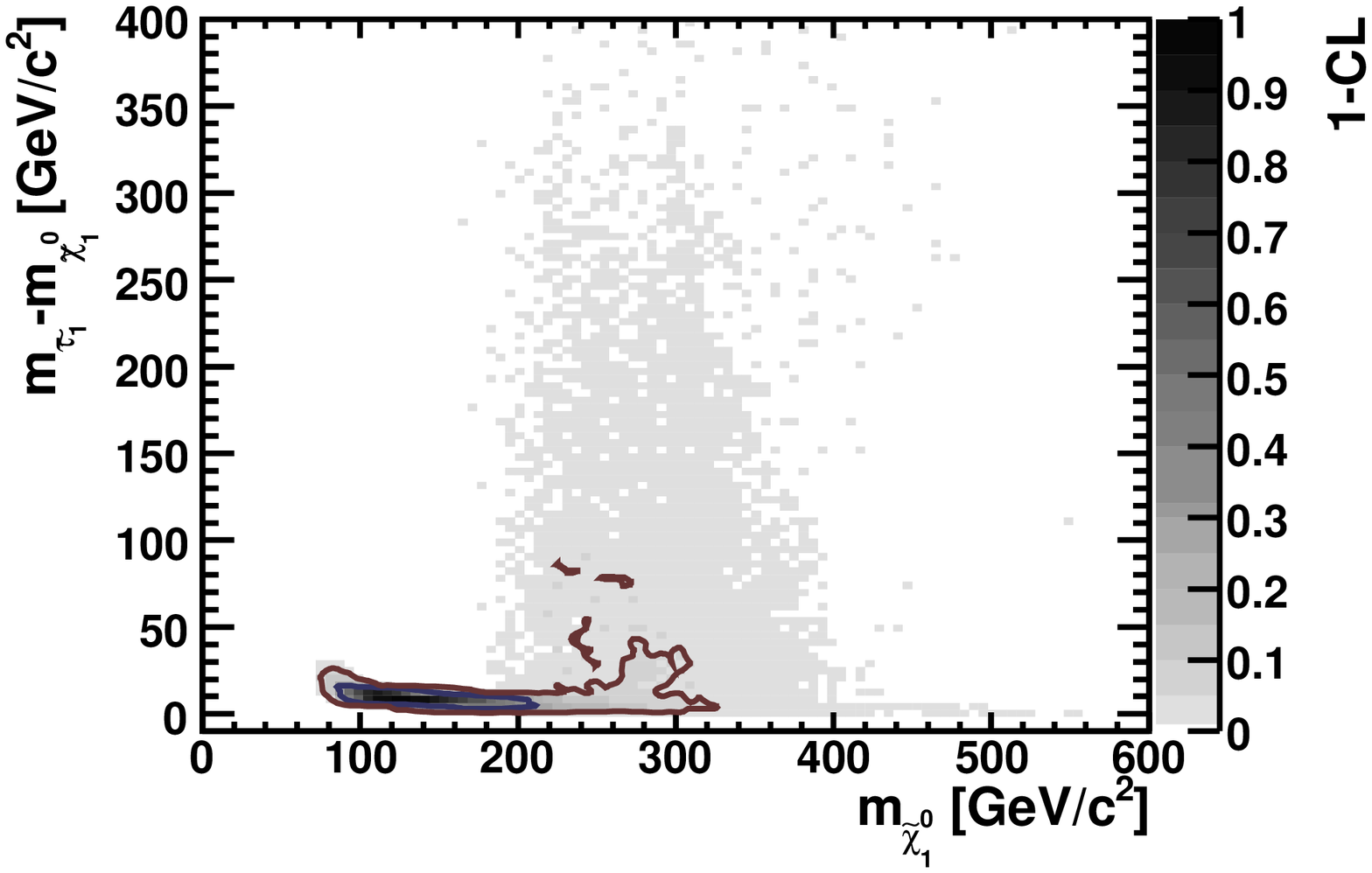}
\hspace{0.2cm}
%\vspace{-0.02cm}
\includegraphics[width=0.45\textwidth]{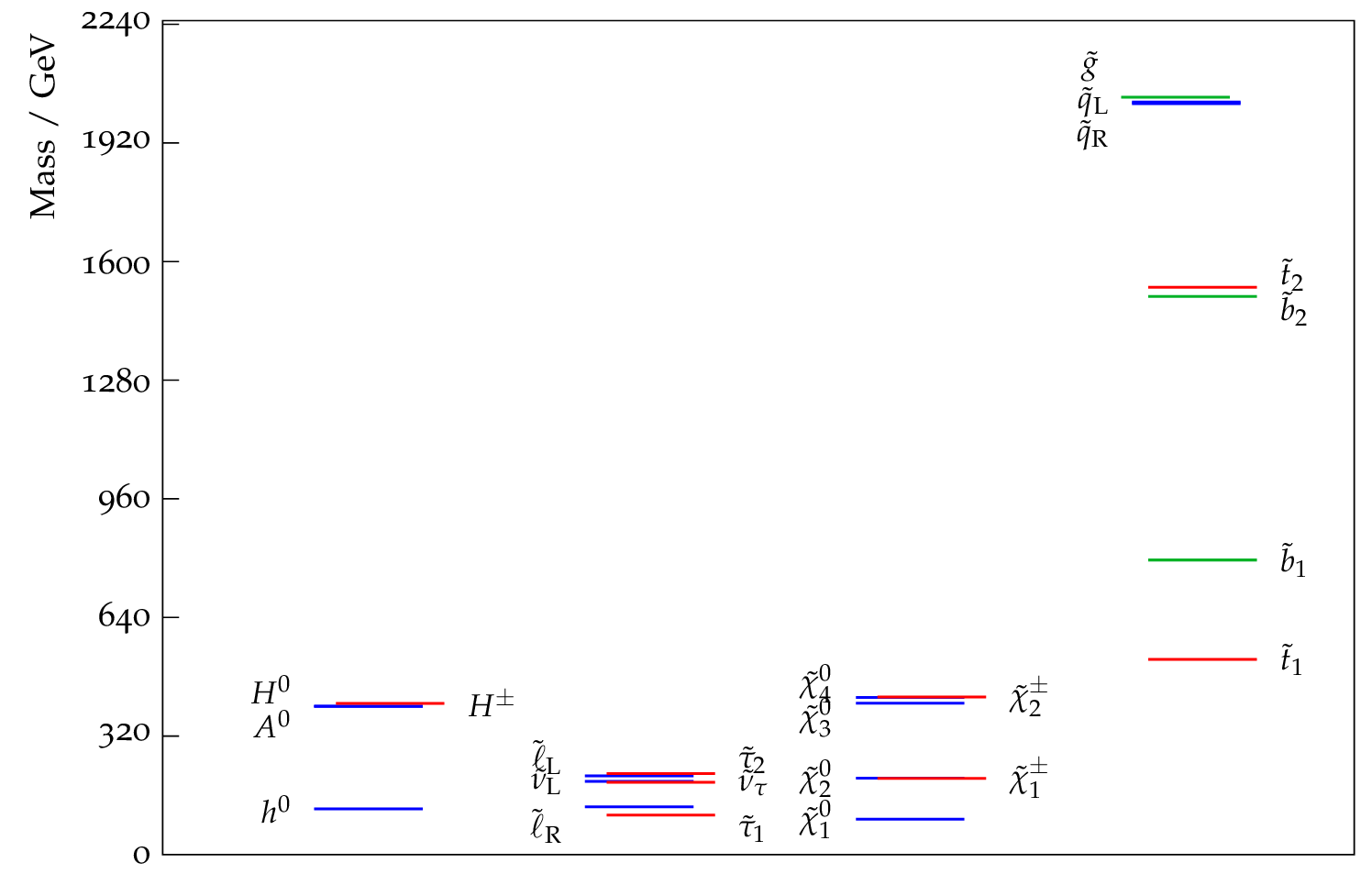}
% \end{center}
  \caption{\label{fig:deltaMstau} Left: Likelihood of a constrained MSSM fit to pre-LHC experimental data in
  the $\Delta M (\ttau, \tz_1) - M_{\tz_1}$ plane. The series of example points discussed here have 
  $\Delta M (\ttau, \tz_1) - M_{\tz_1} = 10$~GeV, right on the strip of highest likelihood, from Ref.~\cite{Buchmueller:2009fn}.  Right: Spectrum of one of the example scenarios, from Ref.~\cite{Baer:2013ula}}.
\end{figure}

Take as an example the regions in parameter space which gained the highest likelihood in fits to 
all pre-LHC experimental data within the constrained MSSM~\cite{Buchmueller:2009fn, Bechtle:2009ty}. 
These fits preferred 
scenarios with a small mass difference of about $10$~GeV between the $\ttau$ NLSP and the 
$\tz_1$ LSP, as illustrated by the likelihood distribution in the left panel of 
Fig.~\ref{fig:deltaMstau}.  Without the 
restriction of mass unification at the GUT scale, the part of the spectrum which is of interest 
to electroweak and flavor precision observables and dark matter, 
{\it i.e.} which is decisive for the fit outcome, is not at all in conflict with LHC results. The right part 
of Fig.~\ref{fig:deltaMstau} shows an example of a spectrum fulfilling all constraints, including
a Higgs boson with SM-like branching ratios at a mass in agreement with the LHC discovery within
the typical theoretical uncertainty of $\pm 3$~GeV on MSSM Higgs mass calculations. The full definition
and further information can be found in Ref.~\cite{Baer:2013ula}.

A series of similar points with increasing top squark mass has been studied recently
with the Snowmass version of the LHC detectors as implemented in Delphes~\cite{Ovyn:2009tx}. If the $\tst_1$
is lighter than $\simeq 500$~GeV, $\tst_1 \tst_1$ production is the dominant mode with a cross-section of a few pb,
followed by electroweakino production, inclusively at the level of $1$~pb at $14$~TeV. Although in this region
of stop and electroweakino masses serious constraints seem to exist in simplified models, a Delphes ``recast" of
the relevant analyses on the non-simplified full model revealed that the sensitivity
of the current data is actually {\itshape not} sufficient. The  study further shows that these points could be 
discovered as a deviation from the SM at LHC14, but only few states could be resolved
and clearly identified~\cite{bib:STC_WP}.

On the other hand, at the ILC running at $\sqrt{s}$ = 500 GeV, 
all sleptons and all charginos and neutralinos can be produced, with sizable
cross-sections - only one of the allowed processes would have
a production cross-section below $1$~fb for both beam polarizations.
The total SUSY cross-section is over $3$~pb in both cases.

\begin{figure}[htb]
  \begin{center}
\includegraphics[width=0.49\textwidth]{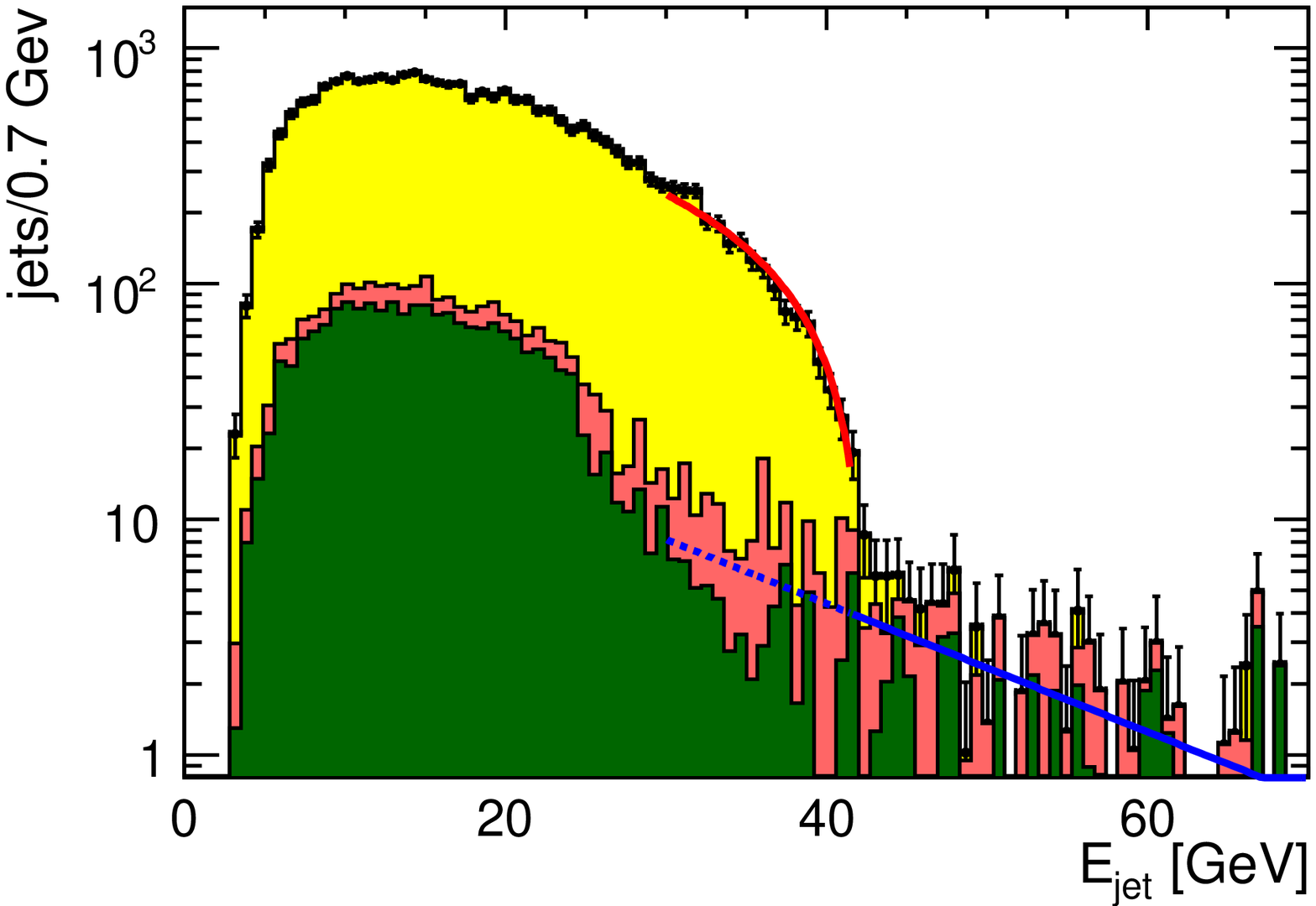}
\includegraphics[width=0.45\textwidth]{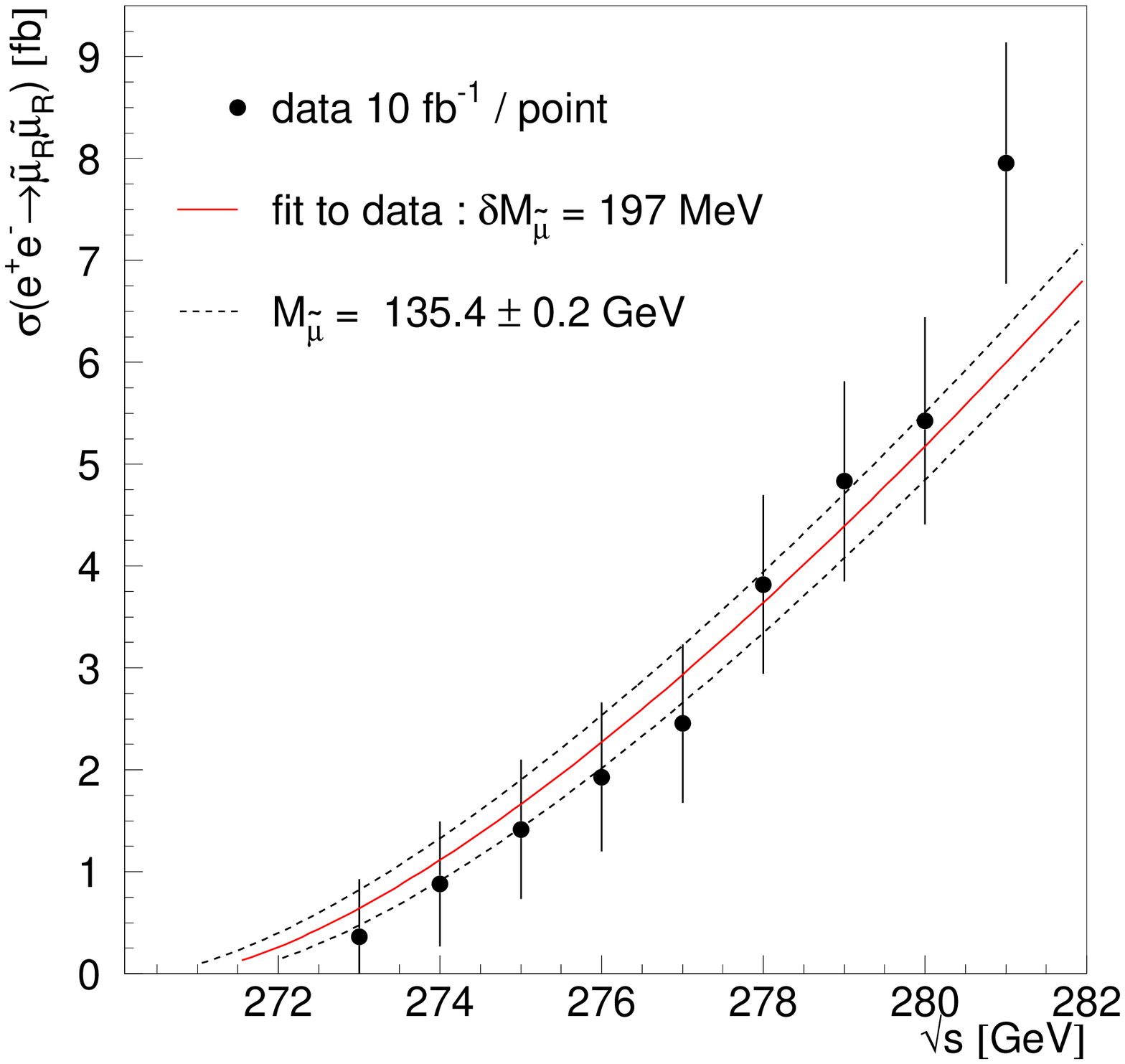}
 \end{center}
  \caption{\label{fig:stc-ilc} Left: \stau{1} spectrum (yellow) and
  background (SM: red, other SUSY: green), with end-point fit. From Ref.~\cite{Bechtle:2009em}. 
  Right: Threshold scan at the $\eeto \smu{R} \smu{R}$ threshold.
   From Ref.~\cite{Brau:2012hv}}.
  %% OR if I can find the plot:
  %% Right: Kinematicly constrained \smu{R} mass-spectrum from \XI0{2} $\to$ \smu{R}$\mu \to \mu \mu \XI0{1}.
  %% From~\cite{...}}.
\end{figure}

In the slepton sector, several of the channels are being studied, or have been
in the past. For example, in \cite{Bechtle:2009em} it was shown that the
\stau{1} mass could be determined to 0.2\% (Fig. \ref{fig:stc-ilc}),
and the \stau{2} mass to 3\% .
Production cross-section for both these modes can be
determined at the level of 4\%, and the polarization
of $\tau$-leptons from the \stau{1} decay could be
measured with an accuracy better than 10\%.
Further studies show that the mass of \smu{R} can be
determined to 0.1\% by a threshold scan \cite{Brau:2012hv}, see Fig. \ref{fig:stc-ilc}.

In the electroweakino sector, 
further constraints on the neutralino and chargino mixing matrices
are possible. Almost all processes
are kinematically accessible at 500 GeV, and all but one (\eeto\XI0{3}\XI0{3}) 
have cross sections above
$1$~fb.
%Although the \stau{1} is the NLSP,
%almost all bosinos have some BR to other final states
%than the notoriously difficult $\tau$-lepton.
The use of beam polarization and tunable $\sqrt{s}$ will further enhance the
power of the observations.
Due to the low background,
cascade decays can be disentangled, and sometimes
fully kinematically constrained.
When that is possible,
one can hope to obtain
even lower uncertainties on the slepton masses,
of the order of 10 MeV \cite{Berggren:2005gp}.
%%   see Fig. \ref{fig:stc-ilc}.

This plethora of precision measurements will allow for precise
determination of SUSY model parameters, and will help to confirm
or rule out different proposed SUSY breaking mechanisms.
In particular, the precision measurement of the LSP and NLSP
properties will connect to cosmology, and answer the question if
SUSY indeed does provide the correct relic density. For more details
on both LHC and ILC capabilities in such models, see~\cite{bib:STC_WP}.

To conclude, this scenario is an excellent case for the LHC-ILC synergy:
LHC will, at the time ILC is commissioned, have discovered the 
stop, and might have some insight into the electroweakino sector. ILC will then
be able to make detailed measurements of masses, couplings and mixings in the
slepton and electroweakino sectors. These refined inputs will help LHC to
interpret the observations in the colored sector, and possibly
hint to new search strategies, including topologies that might previously 
not even
have passed the trigger conditions. In this way,
the results from the two machines, running concurrently, will allow to map out
the full landscape of the model.

\subsection{LHC and ILC complementarity: Electroweakinos }
\label{subsec:ewkinos}
In many SUSY models, colored sparticles are expected to be amongst the heaviest
of the superpartners, and indeed data from LHC8 points to first/second generation 
squarks and gluinos beyond the $\sim 1-1.5$ TeV mass range (at least within the context of simple models 
such as CMSSM).
However, the two chargino and four neutralino states (collectively referred to as {\it electroweakinos})
are expected in most models to be much lighter, and perhaps within kinematic reach of 
both LHC14 and various ILC energy options.

At LHC, the reactions $pp\to \tw_i^+\tw_{i'}^-$, $\tz_j\tz_{j'}$ and $\tw_i^\pm\tz_j$ ($i,\ i'=1,\ 2$ and $j,\ j'=1-4$)
are all expected to occur. Of this panoply of reactions, theoretical projections are that $\tw_1^+\tw_1^-$ 
will be buried beneath $W^+W^-$ and $t\bar{t}$ backgrounds, while $\tz_j\tz_{j'}$ either occur at low rates or are
buried beneath backgrounds. The best hope for LHC seems to be the mixed reaction $pp\to\tw_i^\pm\tz_j$. 
In the case where $|\mu|\gg M_1,\ M_2$ and sfermions are heavy, 
this leads to 1. the clean off-shell trilepton signature from  $\tw_1^\pm\to\ell^\prime\nu_{\ell^\prime}\tz_1$ and 
$\tz_2\to\ell^+\ell^-\tz_1$ decay or 2. to $WZ\to 3\ell$ for $\tw_1^\pm\to W^\pm\tz_1$ and $\tz_2\to Z\tz_1$ or 3. 
to $Wh$ production from $\tw_1^\pm\to W^\pm\tz_1$ and $\tz_2\to h\tz_1$. In the first case, the $m(\ell^+\ell^- )$
invariant mass can be used to yield a high precision measurement of $m_{\tz_2}-m_{\tz_1}$. 
In the case where $|\mu |<M_1,\ M_2$, then wino pair production followed by decay to same sign dibosons 
$pp\to\tw_2^\pm\tz_4\to (W^\pm\tz_{1,2})+(W^\pm\tw_{1}^\mp)$ yields an excellent signature with low backgrounds 
for models with light Higgsinos.

In contrast, at ILC the mixed production reaction isn't allowed due to charge conservation. 
However, the complementary reaction $e^+e^-\to\tw_1^+\tw_1^-$ should be easily seen above background provided that
$\sqrt{s}>2m_{\tw_1^+}$. 
Since the beam energy is precisely known, this allows for extraction of the masses $m_{\tw_1^+}$ and $m_{\tz_1}$ 
to high precision. 
In addition, mixed production $e^+e^-\to \tw_1^\pm\tw_2^\mp$ can allow access to the heavier
$\tw_2^\pm$ chargino state even if $\sqrt{s}<2m_{\tw_2^+}$. Also, the ten $\tz_j\tz_{j'}$ reactions should be easy to spot
provided that $\sqrt{s}>m_{\tz_j}+m_{\tz_{j'}}$. Threshold scans and beam polarization will help to differentiate
these reactions and to discriminate the Higgsino/gaugino components of each $\tz_j$ state which is accessible. 
Even in the case where $e^+e^-\to\tz_1\tz_1$, radiation of initial state photons can be used to tag this reaction
against $e^+e^-\to\nu\bar{\nu}\gamma$ background.
To summarize: while LHC is sensitive to mainly mixed electroweakino production provided 
there are large enough mass gaps between parent and daughter particles, ILC will
be sensitive to a variety of complementary reactions even when small mass gaps occur, such as for
charged and neutral Higgsino pair production. 

A joint LHC/ILC study demonstrates the complementarity of the two machines~\cite{ewkino-wp}.
In this study, a parameter scan is performed in the context of the MSSM
across the three-dimensional parameter space in the $M_1$, $M_2$, and $\mu$ variables,
fixing the Higgs mass to the observed value and $\tan\beta$.
%which does not significantly affect the electroweakino sector.
The masses and branching ratios are obtained from tree-level calculations.
Small mass gaps down to about 200~MeV are studied,
below which the loop corrections become significant.

\begin{figure}
\centering
\includegraphics[width=0.45\textwidth]{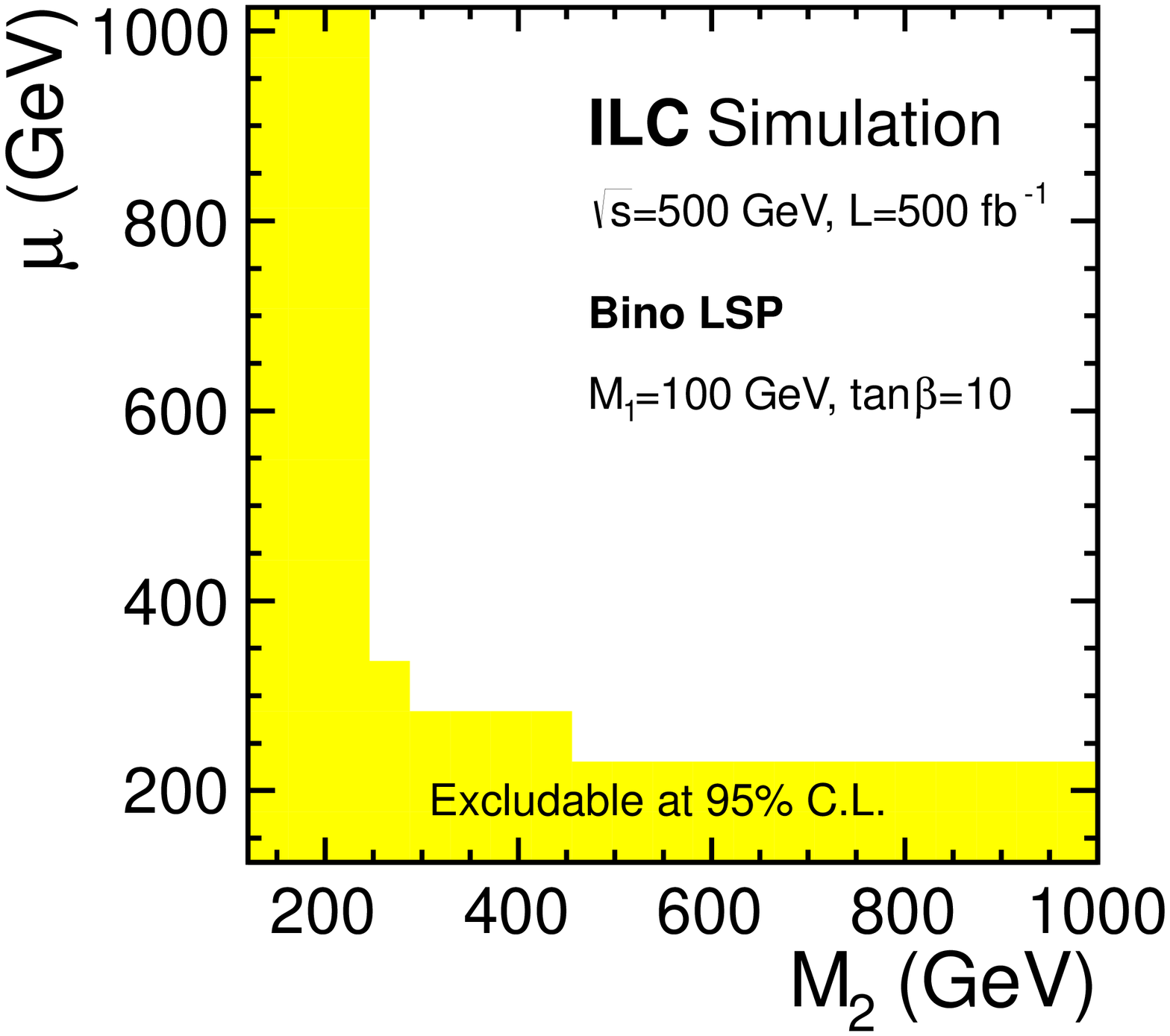}
\includegraphics[width=0.45\textwidth]{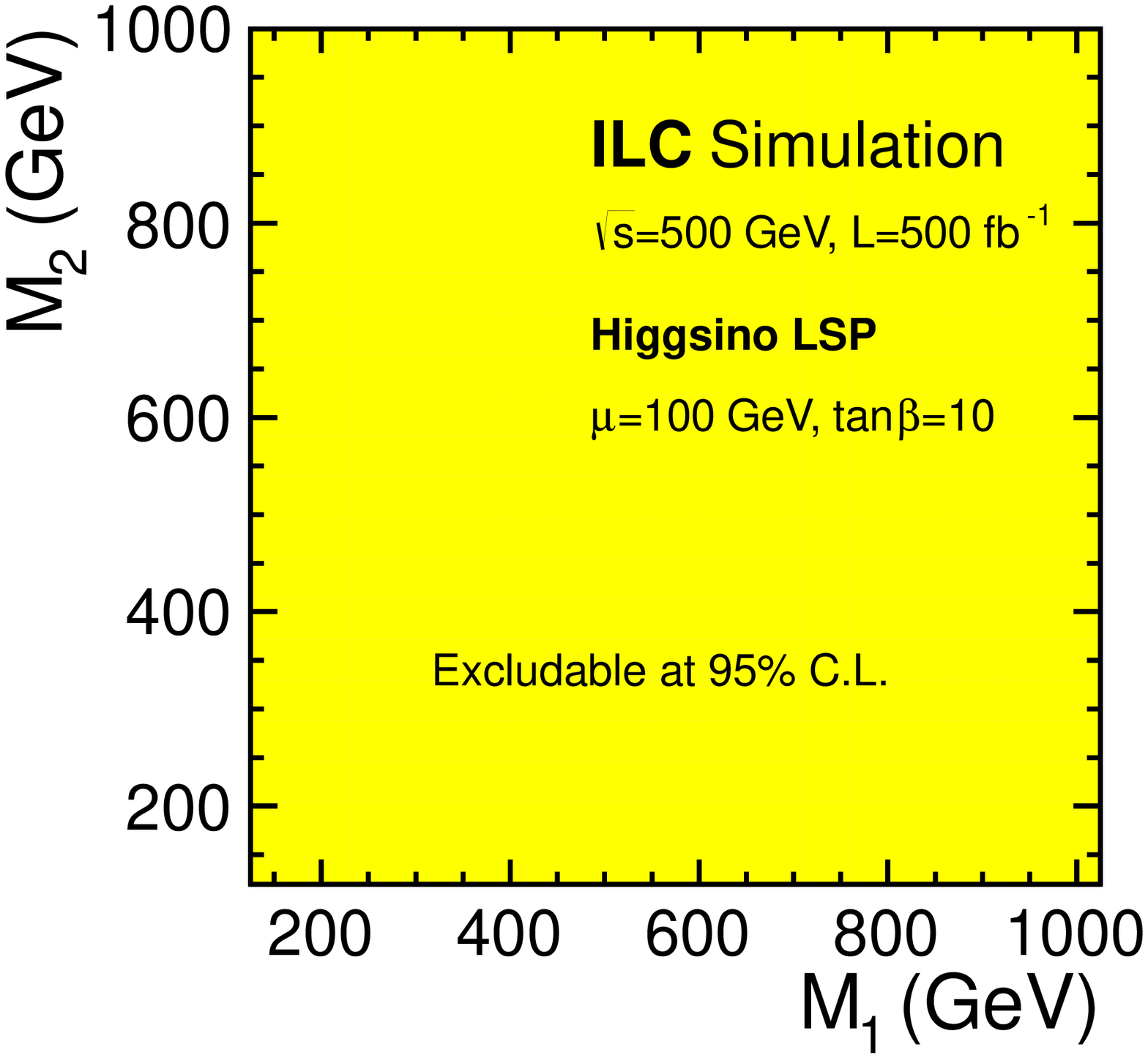}
\caption{Exclusion reach for electroweakinos at the ILC
  studied with fast simulation.
  Left: the case of Bino LSP. Right: the case of Higgsino LSP.
  The shaded region (yellow) is the expected exclusion reach at 95\% confidence level.
}
\label{fig:ewkino}
\end{figure}

For the ILC, two analysis strategies are employed.
First, the signature of four jets plus missing four-momentum
covers the case of medium to large mass differences among the electroweakinos.
Second, the recoil of the initial state radiation photon
is used for the case of small mass differences.
The SGV fast simulation tool~\cite{SGV} is used to simulate the ILC detector response,
which takes into account the effect of particle flow calorimetry.
The expected exclusion reach is shown in Fig.~\ref{fig:ewkino} for the case of the ILC
with $\sqrt{s}=500$~GeV.

In the case of Bino LSP, the ILC is essentially sensitive to
$M_2$ and $\mu$ of up to 250~GeV, which is half the center-of-mass energy;
the NLSP is accessible under this condition and its decays can be detected.
The LHC is expected to be able to cover larger $M_2$ values,
which provide the large mass gaps allowing the detection of the electroweakinos.
The case of LSP pair production without NLSP can be searched
at the ILC with a single photon signature from the ISR
and is covered in Sec.~\ref{subsec:WIMPs}.

In the case of Higgsino LSP with $\mu=100$~GeV,
the LSP and NLSP are accessible regardless of $M_1$ and $M_2$.
The detection of Higgsino decays, which are typically soft,
can be used to exclude the entire parameter space with a light Higgsino,
which is a capability unique to the ILC.

In addition, the ILC can separate the chargino and neutralino contributions in many cases,
providing cross-section and mass measurements at the $O(1)$\% level~\cite{SueharaList,bib:higgsino_paper}.

\subsection{Bilinear R-Parity Violation: Neutrino Physics at Colliders }
\label{subsec:bRPV}
One outstanding question the Standard Model cannot explain is the smallness of neutrino masses.
As explained in Sec.~\ref{sec:intensity}, some of the proposed mechanisms of neutrino mass generation involving supersymmetry can be studied
 at the ILC through precise measurements of properties of new particles. As a concrete example, 
 we will illustrate the ILC's capabilities assuming bilinear $R$-parity violation model 
 of neutrino masses introduced in Sec.~\ref{sec:intensity}.

In this case, the atmospheric neutrino mixing angle can be accessed via the ratio of branching 
ratios of the LSP: 
$BR(\tilde \chi^0_1 \to W \mu)/BR(\tilde \chi^0_1 \to W \tau) \simeq \tan^2\theta_\mathrm{atm}$. 
At the ILC, these branching ratios can be measured from direct LSP pair production even
if all other SUSY particles should be out of kinematic range for direct production. The 
cross section depends on the neutralino mixing and the selectron mass,  but can easily be 
as large as $100$~fb, giving {\it e.g.} 25000 events per year of design ILC operation at 
$\sqrt{s} = 500$~GeV. 

\begin{figure}[htb]
  \begin{center}
\includegraphics[width=0.49\textwidth]{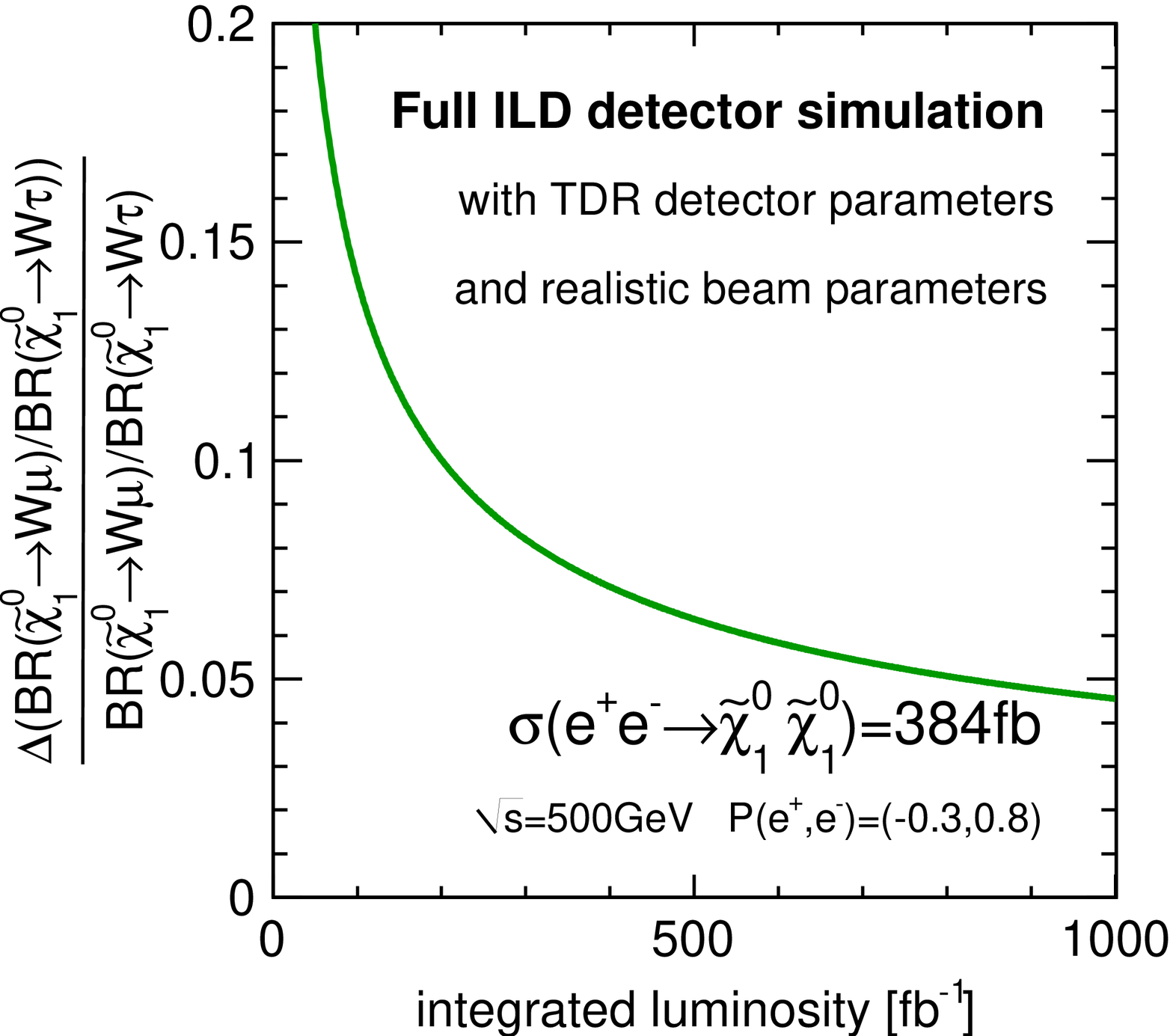}
\includegraphics[width=0.45\textwidth]{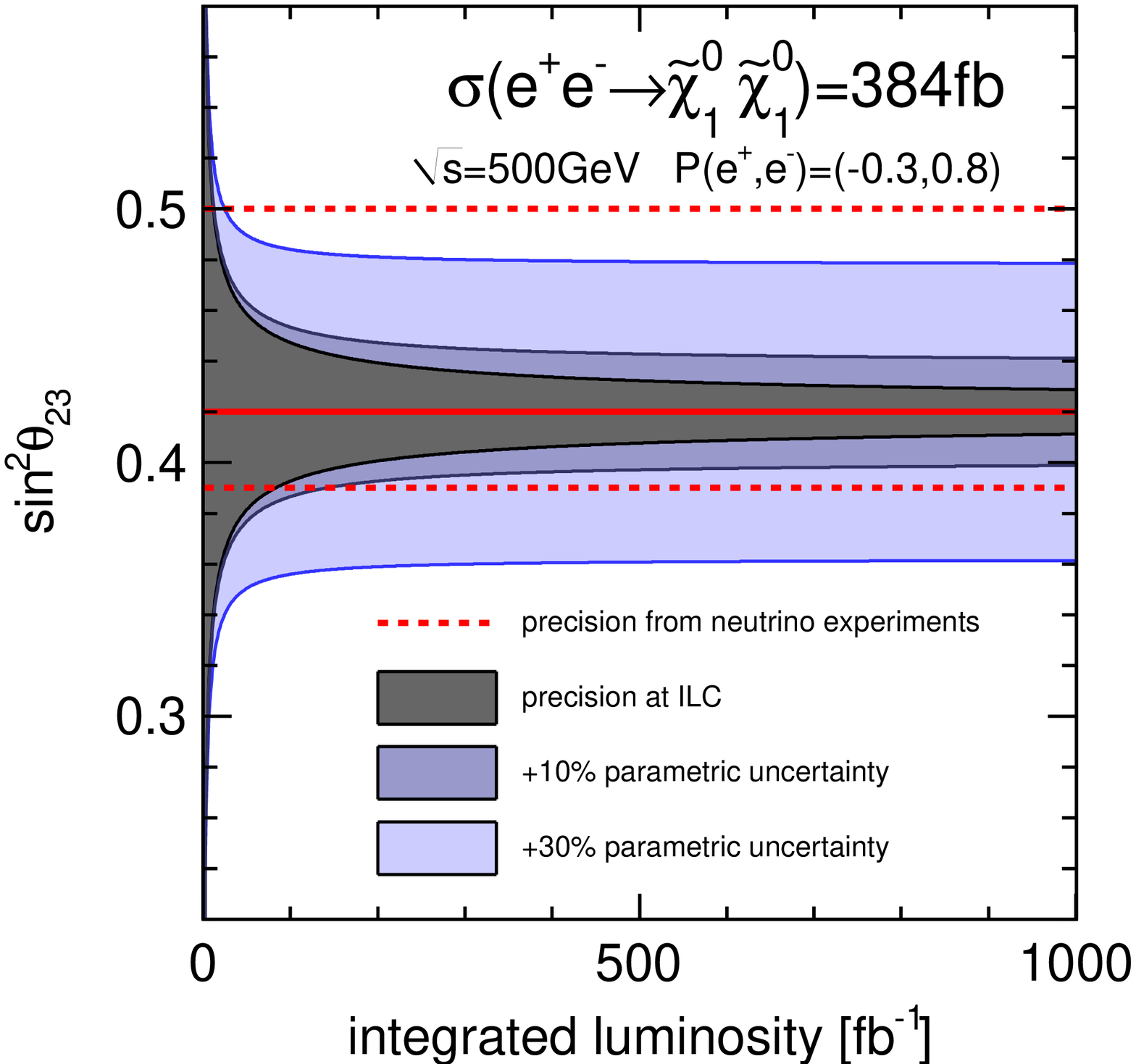}
 \end{center}
  \caption{\label{fig:bRPV} Left: Experimental precision 
  on $BR(\tilde \chi^0_1 \to W \mu)/BR(\tilde \chi^0_1 \to W \tau)$ as a function of the integrated luminosity for the simulated benchmark point. Right: Derived uncertainty on the atmospheric neutrino mixing angle $\sin^2\theta_{23}$, 
  including the additional parametric uncertainty due to limited knowledge of other SUSY parameters.}
\end{figure}

The experimental performance for the branching ratio measurements has been
presented in the ILC TDR~\cite{Intro1} using state-of-the-art full simulation of the ILD detector
concept~\cite{Intro3}. The obtained relative precision on 
$BR(\tilde \chi^0_1 \to W \mu)/BR(\tilde \chi^0_1 \to W \tau)$ is shown in the left panel of 
Fig.~\ref{fig:bRPV}. Already for an integrated luminosity of $250\,$fb$^{-1}$, a relative precision of $7\%$ is 
achieved, which decreases to $5\%$ for $500\,$fb$^{-1}$. The right panel illustrates the 
resulting absolute precision on $\sin^2\theta_\mathrm{atm}$, assuming the current central value
from oscillation measurements, whose precision is indicated by the dashed horizontal lines. 
For the 
collider measurement, the effect of the parametric uncertainty due to limited knowledge of other
SUSY parameters which enter the relation
between the branching ratios and the mixing angle is displayed in addition.

Such a measurement, if in agreement with direct oscillation measurement, could pin down 
the origin of neutrino masses and mixing, which would solve one of the major puzzles in our
current understanding of nature.

\subsection{Production of dark matter in the laboratory? }
\label{subsec:WIMPs}
As discussed in Sec.~\ref{sec:cosmo}, the existence of Dark Matter provides a strong indication 
for the appearance of new phenomena near the electroweak scale. The ILC is a powerful tool for probing the WIMP hypothesis.

\begin{figure}[htb]
\setlength{\unitlength}{1.0cm}
%  \begin{center}
%\includegraphics[width=0.49\textwidth]{WIMPmassres.eps}
%\includegraphics[width=0.45\textwidth]{}
\includegraphics[width=0.62\linewidth]{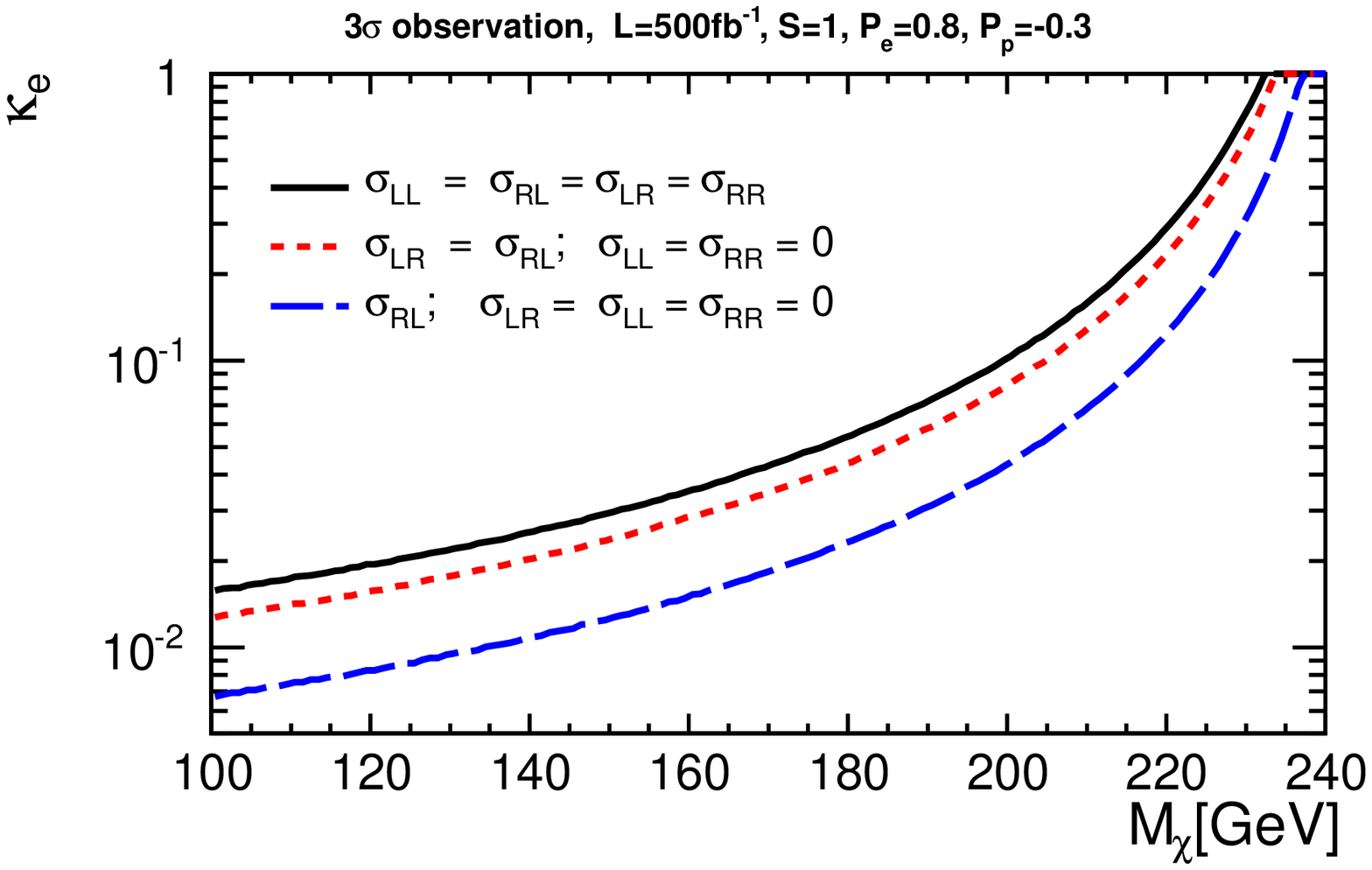}
\hspace{0.1cm}
\includegraphics[width=0.37\textwidth]{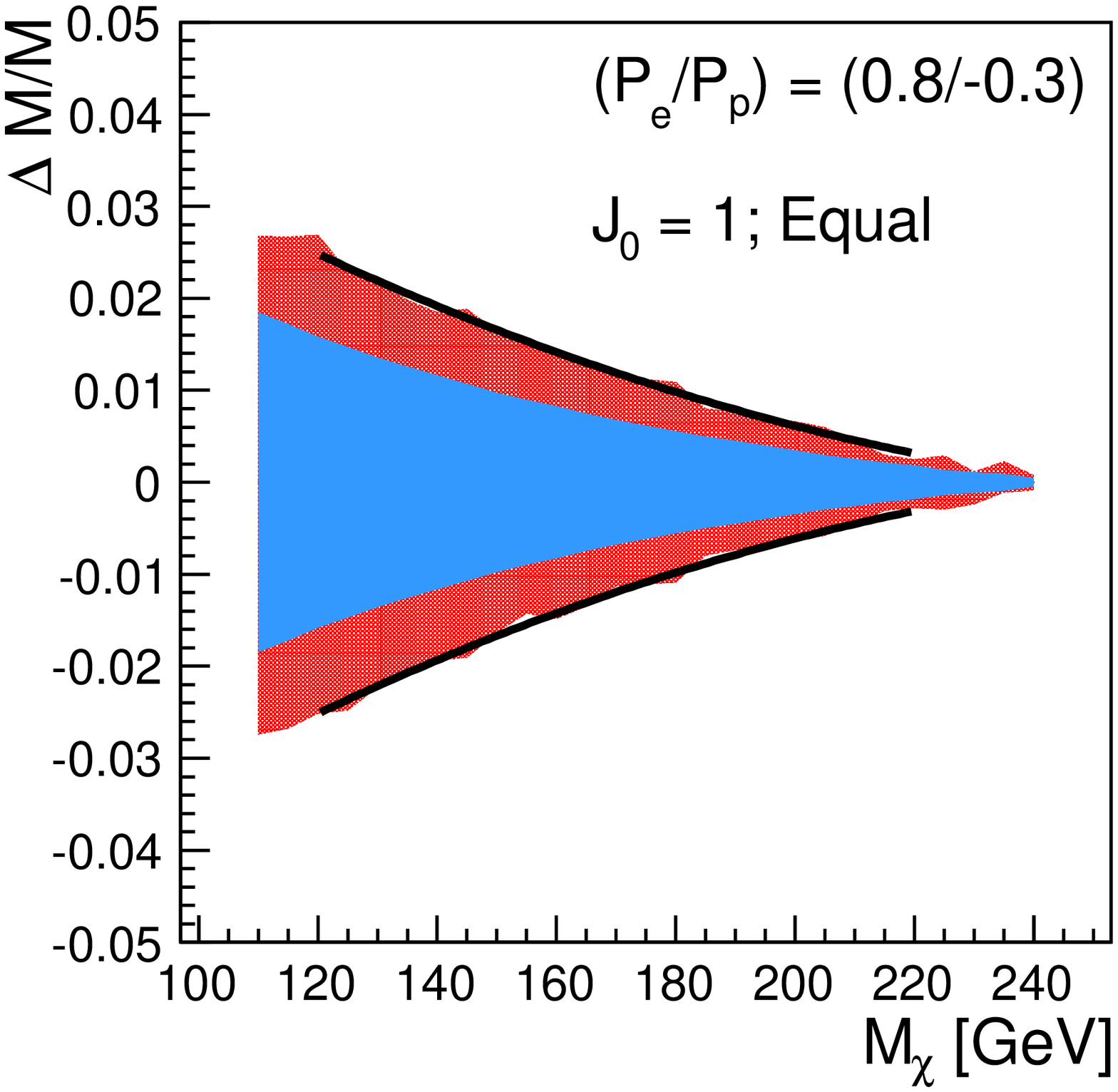}
% \end{center}
  \caption{\label{fig:WIMPs} Left: Observational reach ($3\sigma$) of the ILC for a Spin-1 WIMP in terms of WIMP mass and $\kappa_e$ for three different chiralities of the WIMP-fermion couplings~\cite{bib:chaus} (Note: At WIMP masses lower than $\sim 150\,$GeV, not all assumptions in the relation between the WIMP annihilation and $e^+e^-$ production cross-sections are fully  valid anymore. Signals of low mass WIMPs could still easily be observed, but the interpretation of $\kappa_e$ becomes somewhat less straight forward, since for too low masses the WIMP production is not non-relativistic anymore). Right: Experimental statistical (red) and systematic (blue) precision of the mass reconstruction as a function of the WIMP mass. From Ref.~\cite{Bartels:2012ex}.}
\end{figure}

% \begin{figure}[htb]
% \setlength{\unitlength}{1.0cm}
% %  \begin{center}
% %\includegraphics[width=0.49\textwidth]{WIMPmassres.eps}
% %\includegraphics[width=0.45\textwidth]{}
% \includegraphics[width=0.49\linewidth]{WIMP_kLkR.eps}
% \put(-6.5,1.0){(a) $\kappa_e = \kappa(e^-_Le^+_R)$}
% \includegraphics[width=0.49\linewidth]{WIMP_HP.eps}
% \put(-7.0,1.0){(b) $ \kappa_e /2 = \kappa(e^-_Le^+_R) = \kappa(e^-_Re^+_L)$}
% \vspace{0.01cm}
% \includegraphics[width=0.49\linewidth]{WIMP_kRkL.eps}
% \put(-6.5,4.0){(c) $\kappa_e = \kappa(e^-_Re^+_L)$}
% \hspace{0.34cm}
% \includegraphics[width=0.35\textwidth]{WIMPmassres.eps}
% \put(-0.0,1.0){(d)}
% % \end{center}
%   \caption{\label{fig:WIMPs} (a)-(c) Observational reach ($3\sigma$) of the ILC for a Spin-1 WIMP in terms of WIMP mass and $\kappa_e$ for three different chiralities of the WIMP-fermion couplings. Full line: $P_{e^-} = P_{e^+} = 0$, dotted line: $P_{e^-} = 0.8, P_{e^+} = 0$, dashed line: $P_{e^-} = 0.8, P_{e^+} = 0.6$. From Ref.~\cite{Bartels:2009fa}. (d) Experimental statistical (red) and systematic (blue) precision 
%   of the mass reconstruction as a function of the WIMP mass. From Ref.~\cite{Bartels:2012ex}.}
% \end{figure}
% 
If the WIMPs have a non-negligible coupling to electrons, they can be pair produced in electron-positron 
collisions: $e^+e^- \ra \chi \chi$. The WIMPs themselves leave a collider detector without a trace.
Nevertheless such events can be identified when the WIMP system recoils against an energetic photon
from initial state radiation, very similar to the mono-jet and mono-photon signatures searched for
at the LHC~\cite{bib:ATLAS_mono, bib:CMS_mono}. In case of the ILC, the potential of the mono-photon signature
has been studied at theoretical level~\cite{Baer:2001ia,Birkedal:2004xn,Dreiner:2012xm,Chae:2012bq}, as well as in full 
detector simulation of the ILD detector concept~\cite{Bartels:2009fa,Bartels:2012ex}. The special environment at the ILC, 
with a fully controllable initial state of the collisions, including the beam polarization, and the backgrounds at 
typical electroweak level, enables sensitivities to cross-sections of the order of a few fb, depending on the
details of WIMP properties and assumed operation mode of the ILC. This means that the ILC can discover this signature 
even if annihilation to electrons provides only a small fraction $\kappa_e$ of the total dark matter annihilation rate in the early universe. This is illustrated in Fig.~\ref{fig:WIMPs}(a)-(c), which show the $3\sigma$ observation reach for an integrated luminosity of $500$~fb$^{-1}$ at a center-of-mass energy of $500$~GeV for a spin-1 WIMP and three different assumptions on
the helicity structure of the WIMP-electron coupling~\cite{Bartels:2009fa}. Beam polarization increases the reach in $\kappa_e$
by about an order of magnitude.

If a signal is observed, the shape of 
the photon energy spectrum is a powerful tool to pin down the properties of the WIMP candidate. As an example, Fig.~\ref{fig:WIMPs}(d) shows the achievable mass resolution of a few percent as a function of 
the WIMP mass as obtained in full detector simulation, including both statistical and systematic uncertainties.
The statistical uncertainties are based on an integrated luminosity of $500$~fb$^{-1}$ and an unpolarized production cross-section of $100$~fb$^{-1}$, which  for a WIMP mass of $150$~GeV corresponds to $\kappa_e \simeq 0.5$ in case of a spin-1/2 WIMP and to $\kappa_e \simeq 0.25$ for a spin-1 WIMP. For more details, see~\cite{Bartels:2012ex}. 
 The measurement of the polarized cross-sections for all four possible beam polarization configurations will reveal
the helicity properties of the WIMP's interaction with electrons, and the shape of the photon energy
spectrum can be used to determine the dominant partial wave of the WIMP pair production process. 

If the WIMP-electron 
interaction is mediated by a heavy particle with mass well above the WIMP mass, the interaction can be described in
an effective operator language, {\it e.g.} as a 4-fermion operator for a spin-1/2 WIMP~\cite{Dreiner:2012xm,Chae:2012bq}. 
In this case, ILC measurements can determine the structure of the effective operator(s), providing indirect information on the 
nature of the heavy mediator. 

Currently, the LHC as well as direct detection experiments are searching for WIMPs via their interactions with quarks and gluons. 
The ILC will provide a complementary approach, relying on the WIMP-electron interaction instead. Since the couplings of WIMPs
are very model dependent, such complementarity is of great value. Indirect detection experiments can search for photons and
positrons produced in WIMP annihilations mediated by WIMP-electron interactions, but are primarily sensitive to WIMP masses 
in the few hundred GeV region and above, while the ILC is sensitive to lower mass WIMPs. 
For a more detailed discussion of complementarity between direct, indirect, and collider searches for WIMPs, 
see {\it e.g.} Refs.~\cite{Baer:2004qq,Bauer:2013ihz}.

%\paragraph{with somewhat lower priority:}
%
%\subsection{ADD (???)}
%\label{subsec:ADD}
%\input{ADD}

\subsection{Little Higgs: an alternative to SUSY}
\label{subsec:littlehiggs}
Many alternatives to supersymmetry have been proposed as candidates for physics at the weak scale. While the discovery of the light Higgs eliminates some of these alternatives, many viable options remain. Just as in the case of supersymmetry, the ILC can play a vital role in exploring the physics of these models. As an example, consider the Littlest Higgs model with T-parity, based on $SU(5)/SO(5)$ symmetry breaking pattern~\cite{LHT}. This model is a useful benchmark because its phenomenology has been studied extensively. The parameter space of the model is constrained by precision electroweak constraints~\cite{PEW,FTP}, as well as direct searches for the fermionic top partner~\cite{FTP,TPexp} and T-odd partners of quarks~\cite{Todd} at the LHC. Nevertheless, viable parameter space remains where weakly-coupled new particles could be within the kinematic range of the ILC; in fact, this is precisely the part of the parameter space preferred by naturalness considerations. In particular, the lightest T-odd particle (LTP)-- typically the partner of the photon and an attractive dark matter candidate~\cite{LHDM}-- is likely to be accessible. T-odd partners of the electroweak gauge bosons, as well as partners of charged leptons and neutrinos, could also be directly produced at the ILC. If this is the case, the ILC would be able to pursue a program of precise determination of the masses and couplings of all kinematically accessible states. This would in turn provide a quantitative test of the structure of the Little Higgs model, as well as allow for a precise prediction of the LTP relic density which can then be compared with the observed cosmological abundance of dark matter. 

\begin{figure}[htb]
  \begin{center}
\includegraphics[width=0.49\textwidth]{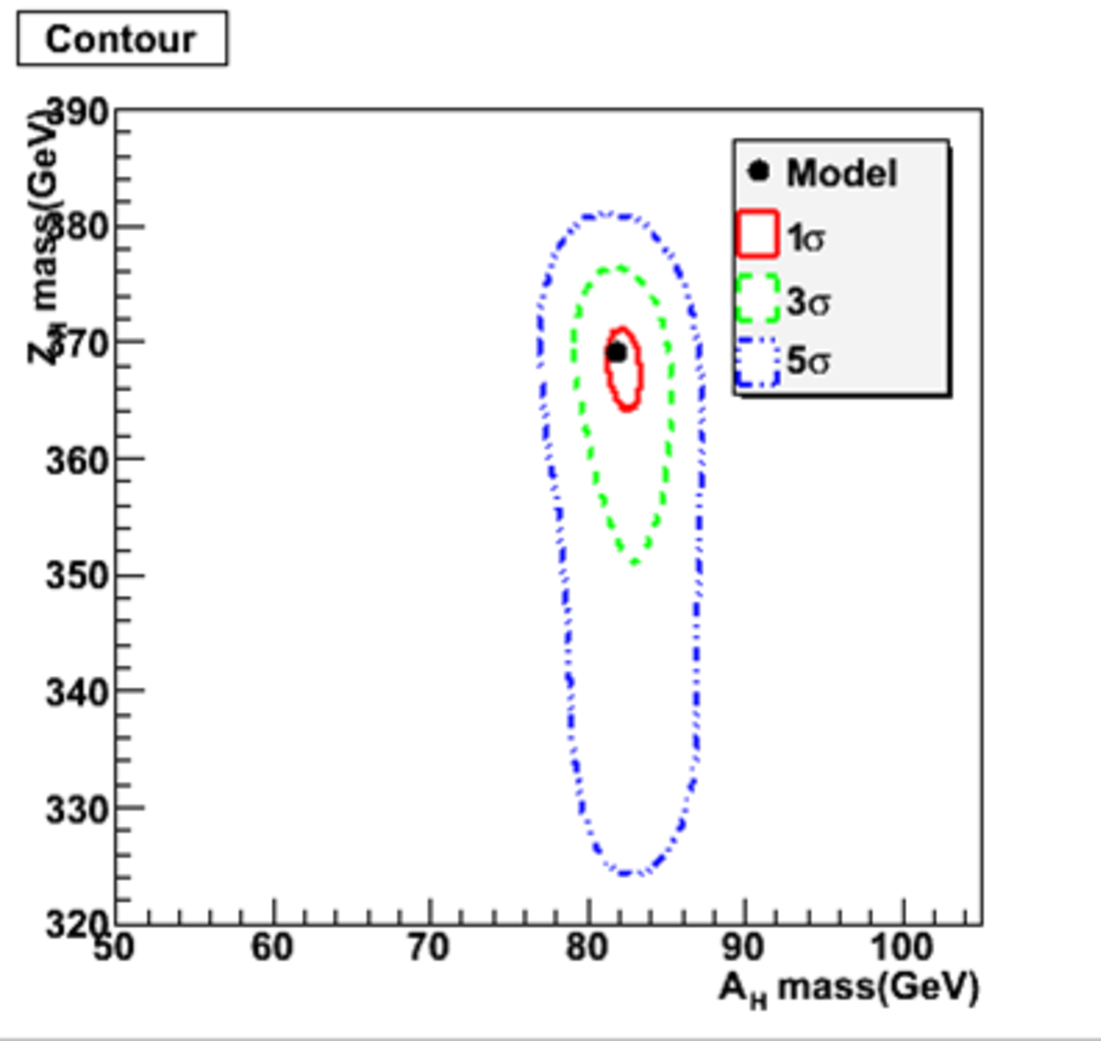}
\includegraphics[width=0.49\textwidth]{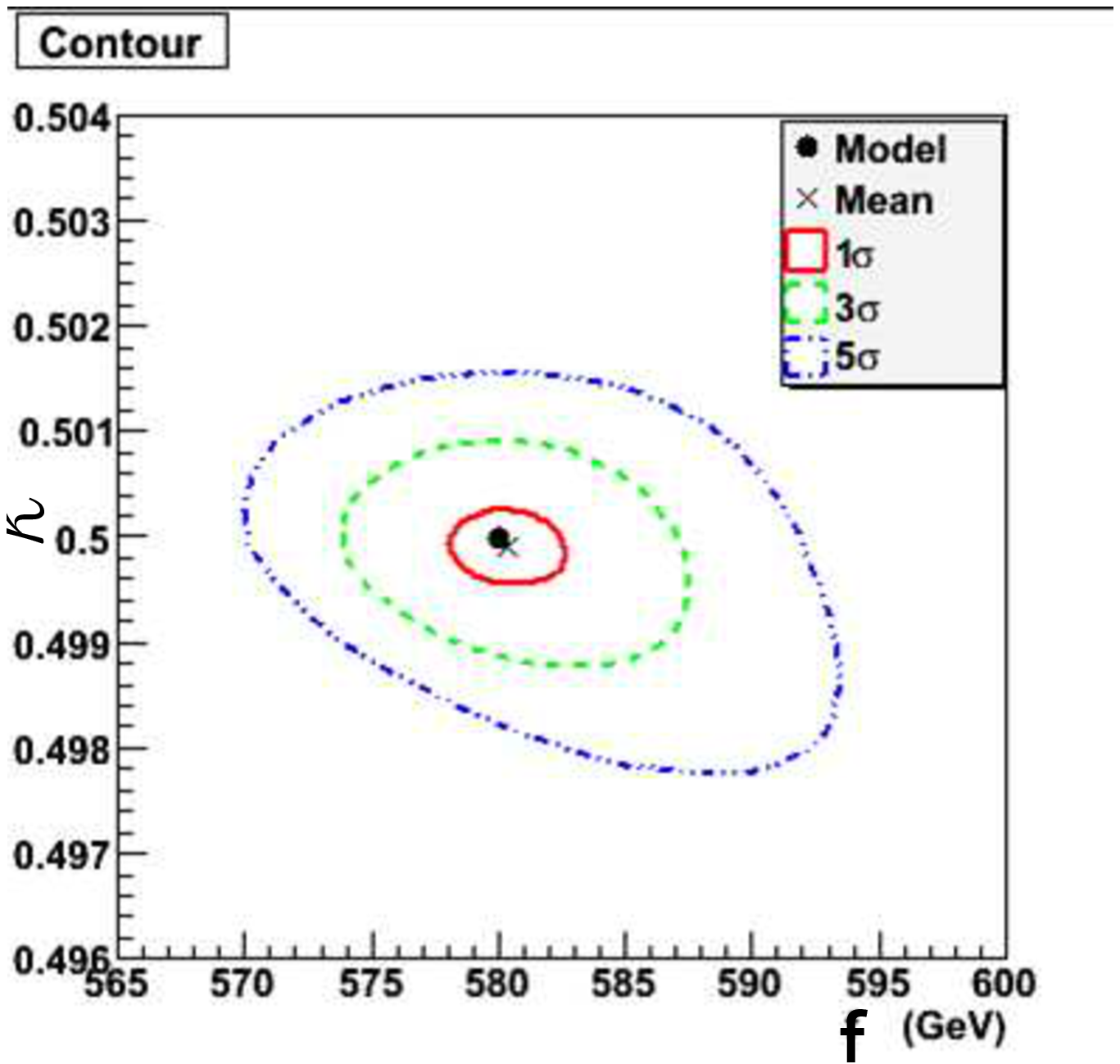}
 \end{center}
  \caption{\label{fig:Asano} Accuracy of T-odd particle mass (left) and model parameter (right) determination for a representative benchmark point in the parameter space of the Littlest Higgs model with T-parity. From~\cite{Asano}.}
\end{figure}

A quantitative study illustrating these points has been presented in Ref.~\cite{Asano}. This study assumed that the LHT model is realized in nature, with the non-linear sigma model scale $f=580$ GeV, and T-odd lepton Yukawa coupling $\kappa=0.5$. For detailed description of the structure of the model and notation, see Refs.~\cite{PEW,LHDM}. For these parameters, the LTP is the T-odd ``heavy photon" $A_H$, with the mass of $82$ GeV. Other low-lying T-odd particles include the heavy $W$ and $Z$ bosons, $W_H$ and $Z_H$, with masses about $370$ GeV; and charged lepton and neutrino partners, $e_H$ and $\nu_H$, at  about $400$ GeV. The Higgs mass of $m_H=134$ GeV was assumed.\footnote{Since Ref.~\cite{Asano} appeared, LHC data provided a slightly lower value of $m_H$. In addition, the charged T-odd lepton mass assumed in~\cite{Asano} might be inconsistent with the lack of excess events in the searches for direct slepton pair-production~\cite{LHC_slepton}, although a careful re-analysis of the data in this context would be required to establish this. In any case, the parameter point used in~\cite{Asano} is sufficiently close to viable regions of the parameter space to provide a useful illustration of the ILC capabilities.} Monte Carlo studies (including fast ILD detector simulation with JSFQuickSimulator package) were performed for the process $e^+e^-\to A_H Z_H$ at 
$\sqrt{s}=500$ GeV, and for the processes $e^+e^-\to Z_HZ_H, W_H^+W_H^-, e_H^+e_H^-$, and $\nu_H\nu_H$ at  $\sqrt{s}=1$ TeV, with integrated luminosity of $500$ fb$^{-1}$ in each case. The main results of the analysis are summarized in Fig.~\ref{fig:Asano}. The accuracy of mass determination for the heavy gauge bosons $A_H$ and $Z_H$ is at or below 1\% level. Precise determination of the LTP mass is especially interesting, since it provides a prediction of its relic abundance. The dominant LTP annihilation channel in this model is $A_HA_H\to W^+W^-$ via an $s$-channel Higgs exchange~\cite{LHDM}. Since the $A_H$ coupling to the Higgs is related to the SM gauge coupling $g^\prime$, and the Higgs mass is already known, the mass of $A_H$ is the only missing ingredient in evaluating its relic density. In addition, combining the measurements of T-odd gauge boson and lepton production provides a precise determination of the LHT mass scale $f$ and the T-odd lepton Yukawa $\kappa$, at a per-mille or better accuracy. An overall consistency of the data with the predictions of the LHT model at this level will provide a sensitive test of the structure of the model.

%
% ****************************************************************************

\section{Summary}
\label{sec:summary}
With the publication of the ILC TDR~\cite{Intro1}, the blueprint is available 
to begin construction of the International Linear Collider. 
Our goal in this white paper was to lay out briefly and succinctly the physics case 
for the ILC project from the perspective of Physics Beyond the Standard Model.
%We urge the US community to jump on board, join with Japan and Europe, 
%and to provide the critical momentum needed to push the ILC from concept to reality.

The case from SM physics is clear: the ILC at $\sqrt{s} \geq 250$ GeV would be a
Higgs factory, allowing for precision determination of virtually all properties of
the newly discovered 125 GeV resonance by LHC once the center-of-mass energy is raised to
above the $t\bar{t}h$ and $Zhh$ thresholds. In particular, we would like to know:
is the new resonance a SM Higgs boson, or do its properties point to 
new physics, and a revision of the laws of nature as we know them?

%From the perspective of BSM, the case is equally clear, and perhaps even more compelling.
From the perspective of BSM the argument is perhaps more subtle, but still  compelling.
Even if LHC running at $\sqrt{s}\sim 13-14$ TeV does not discover any further new particles, the ILC will remain a discovery machine, covering a variety of scenarios that present difficulties for the LHC. Some of these scenarios were listed in the Introduction. For example, in the context of supersymmetry, the ILC has a unique discovery capacity for spectra with small $m_\mathrm{NLSP}-m_\mathrm{LSP}$ mass gaps. Such spectra are
highly motivated by SUSY naturalness (with a set of highly compressed light Higgsino states) or
by dark matter co-annihilation scenarios.

In the case where LHC does discover new particles, then ILC will likely function as a precision
microscope, determining the properties of low-lying new particles using its unique capabilities of
low background, known beam features, cleanliness of events, capacity for threshold scans and polarizable beams.

We presented several case studies, or stories, to illustrate our case. For example:
\begin{itemize}
\item {\it Concluding the story of SUSY naturalness:} For SUSY to naturally accommodate the weak scale-- {\it e.g.}
$m_Z\simeq 91.2$ GeV and $m_h\simeq 125$ GeV-- a necessary condition is the presence of light Higgsino states
in the $100-300$ GeV range, the lower the better. The compressed mass spectra of the Higgsinos make them difficult 
to detect at LHC, and projections indicate that {\it natural SUSY} can elude LHC searches.
Since the light Higgsinos-- even with a small mass gap -- are easily seen in the clean ILC environment, the ILC 
will either become a Higgsino factory, or rule out the idea of SUSY naturalness.
\item {\it SUSY is simplified at ILC:} As the energy of ILC is increased, it may surpass 
threshold for pair production of low-lying new particles. In the case of SUSY, 
the production and decay of NLSP should be very simple, 
allowing for high precision determination of the new particle's properties.
\item {\it SUSY is complex:} As $\sqrt{s}$ is raised even further, then at least in the case of SUSY
more and more new states should become accessible to direct production. The new particle
cascade decays can become highly complex: in many cases, the largest background to SUSY comes from other SUSY
processes. The high precision environment of ILC, along with information gleaned from lower energy studies of
NLSP properties, will allow experimenters to disentangle the complicated cases arising from 
production of multiple new states of matter.
\item {\it LHC and ILC complementarity:} In the realm of electroweakino pair production, it could be the case
that LHC sees a signal of mixed chargino-neutralino production followed by decay to trileptons, or possibly
production of heavy wino pairs followed by decays to same-sign dibosons in the case where Higgsinos are light.
In these cases, ILC would produce the complementary reactions of chargino pair production (usually buried beneath $WW$ 
and $t\bar{t}$ background at LHC) or mixed neutralino production (which is also difficult-to-impossible at LHC).
In addition, while some mass gaps are measurable at LHC to high precision, ILC will be able to make model-independent
individual sparticle mass measurements owing to the unique kinematics of $e^+e^-$ collisions.
\end{itemize}

In all these cases-- and in many others which are not discussed here owing to brevity requirements-- 
it is clear that {\it the ILC is the precision instrument of choice} not only for discovery of new 
fundamental particles, but also for precision characterization of their quantum numbers and
other properties. Such measurements will have a profound impact on our understanding of the physical laws at the fundamental level, 
and upon our understanding of the cosmos and its evolution from the Big Bang up to the present day. 
%We strongly urge our colleagues-- scientists, government officials and taxpayers--
%to support the ILC project so that construction may begin as soon as possible.

%\section{Acknowledgments}

%\section{Bibliography}
% ****************************************************************************
% BIBLIOGRAPHY AREA
% ****************************************************************************

\bibliographystyle{unsrt}
\begin{footnotesize}
% IF YOU DO NOT USE BIBTEX, USE THE FOLLOWING SAMPLE SCHEME FOR THE REFERENCES
% ----------------------------------------------------------------------------

% ----------------------------------------------------------------------------

\end{footnotesize}

% ****************************************************************************
% END OF BIBLIOGRAPHY AREA
% ****************************************************************************

\end{document}